\documentclass{appolb}
\usepackage{amsmath}
\usepackage{slashed}
\usepackage{graphicx}
\usepackage{empheq}
\newcommand*\widefbox[1]{\fbox{\hspace{2em}#1\hspace{2em}}}

\begin{document}
\eqsec  
\title{Notes on Anomaly Induced Transport%
\thanks{Based on lectures presented at 56. Cracow School on Theoretical Physics, 
 May 24 - June 1 2016, Zakopane, Poland and the APCTP focus workshop on 
 Holography and Topology of Quantum Matter, August 22 - August 29, 2016, Pohang, Korea.}%
}
\author{Karl Landsteiner
\address{Instituto de F\'{\i}sica Teórica UAM/CSIC, C/ Nicolás Cabrera
13-15,\\
Universidad Autónoma de Madrid, Cantoblanco, 28049 Madrid, Spain}
\\
}
\maketitle
\begin{abstract}
Chiral anomalies give rise to dissipationless transport phenomena such as 
the chiral magnetic and
vortical effects. In these notes I review the theory from a quantum field 
theoretic, hydrodynamic
 and holographic perspective.
A physical interpretation of the otherwise somewhat obscure concepts of {\em 
consistent} and
{\em covariant} anomalies will be given. Vanishing of the CME in strict 
equilibrium will be connected
to the boundary conditions in momentum space imposed by the regularization. The 
role of the gravitational
anomaly will be explained. That it contributes to transport in an unexpectedly 
low order in the
derivative expansion can be easiest understood via holography. 
Anomalous transport is supposed to play also a key  role in understanding the 
electronics of advanced materials,
the Dirac- and Weyl (semi)metals. Anomaly related phenomena such as negative 
magnetoresistivity, anomalous
Hall effect,  thermal anomalous Hall effect and Fermi arcs can be understood via 
anomalous transport. 
Finally I briefly review a holographic model of Weyl semimetal which allows to 
infer a new phenomenon
related to the gravitational anomaly: the presence of odd viscosity.  
\end{abstract}
  
\section{Introduction}
Symmetries are one of the most fundamental concepts of modern physics. The same 
is true for 
quantum theory. However, sometimes these two are in conflict with each other. 
More precisely a symmetry
present on the level of classical Lagrangian might not be compatible with quantum 
theory. When this
happens we speak of a quantum anomaly 
\cite{Bell:1969ts,Adler:1969gk,Bertlmann:1996xk}. 

What shall concern us here specifically are chiral anomalies. These are 
intimately related to the fact that
in even space-time dimensions the Lorentz group has two unitarily inequivalent 
spinor representations giving
rise to left- and right-handed spinors. For massless fermions independent phase 
rotations of left- and
right-handed spinors are symmetries of the classical theory. On the 
quantum level at best one linear combination
of these two symmetries can be preserved. 

In the realm of high energy physics the prime example of a physical phenomenon 
induced by the incompatibility of 
chiral symmetries with quantum theory is the decay of the neutral pion into two 
photons. Besides explaining
such otherwise forbidden (or strongly suppressed) processes in particle physics 
anomalies also place very
stringent consistency conditions on gauge theories. Gauging an anomalous 
symmetry leads to violation of 
unitarity. The divergence of the current couples to the longitudinal
gauge degrees which normally corresponds to zero norm states. Anomalies lead to 
scattering of 
physical states into zero norm states and therefore destroy unitarity. 
Alternatively one can allow a mass term for the gauge field, then however 
renormalizability is 
lost \cite{preskill1991gauge}. 
Even when the symmetries are not gauged anomalies do place very stringent 
conditions on the strong dynamics
of gauge theories. 't-Hooft \cite{'tHooft:1979bh} argued that the spectrum of 
chiral fermions in a gauge theory is protected by
this type of anomalies appearing in global symmetries. These constraints of 
``anomaly matching'' between 
(weakly coupled) high energy theories and (strongly coupled) low energy 
effective theories can be exploited to
get a handle on otherwise difficult to understand strong gauge dynamics. The 
power of anomalies lies in the fact
that they are subject to a non-renormalization theorem \cite{Adler:2004qt} 
stating that the anomaly is exact as an operator relation at one loop.

In the recent years anomalies have also emerged as the leading concept 
that allows to understand 
(and discover) unusual transport phenomena of quantum many body physics 
involving chiral fermions. 
In high energy physics this is relevant to the physics of the quark gluon plasma 
as created in heavy ion
collisions at RHIC and LHC. Anomalies have been invoked to predict charge 
asymmetries in the final state
of a heavy ion collision \cite{Kharzeev:2007jp,Fukushima:2008xe,Burnier:2011bf}  
and indeed charge 
asymmetries consistent with the prediction of anomalous transport
theory have been detected in experiments at RHIC and LHC 
\cite{Adamczyk:2015eqo,Belmont:2014lta}. 
In astrophysics anomalous transport phenomena have been suggested to explain the 
sudden acceleration suffered
by neutron stars at birth (neutron star kicks) \cite{Kaminski:2014jda} in 
cosmology as origin of primordial magnetic fields
\cite{Giovannini:1997eg}. 

But (and probably somewhat surprisingly) anomalous transport phenomena are about 
to play also a lead role
in condensed matter physics. It is already well established that quantum Hall 
physics (see \cite{Tong:2016kpv} for a recent review) 
can be described in a quantum field theory language via anomaly inflow 
\cite{Callan:1984sa} from bulk to boundary of a 
topologically nontrivial insulator. 
More recently also the bulk physics of three (space) dimensional metals has been 
argued to be
governed by chiral anomalies, e.g. via the phenomenon of negative 
magnetoresistivity.  Of course these
are not ordinary metals but very special ones in which the Fermi surface lies at 
or very near linear band touching
points \cite{Volovik:2003fe,Vafek:2013mpa,WSMreviewTV}. In these cases the 
effective low energy electronic excitations near the 
band touching points are chiral fermions
and the theory of anomalous transport can be applied to infer and describe a 
variety of exotic transport phenomena.

The aim of these lectures is to give an introduction to the subject 
with emphasis on making the underlying
quantum field theoretical concepts as clear as possible. If one understands as a 
quantum field theory a prescription
of how to compute correlation functions of (gauge invariant) operators then 
string theory derived holography needs also
to be taken into account. Indeed holography has played a major role in the 
modern area of anomalous transport and
many subtleties arising when dealing with anomalies are most 
easily understood using the holographic
framework 
\cite{Newman:2005hd,Banerjee:2008th,Erdmenger:2008rm,Yee:2009vw,Gynther:2010ed,Landsteiner:2011iq}. 

These notes are organized as follows: in section 2 we will review chiral 
triangle anomalies. Particular emphasis
will be made on the ambiguities  in the regularization procedure and 
how they can be fixed by physical constraints. 
This will lead to the
concepts of {\em covariant} and {\em consistent} anomalies 
\cite{Bardeen:1984pm}.  
In section two we will discuss the Landau level quantization of chiral fermions. 
We will see
how the (covariant) anomaly arises as a conflict between normal-ordering and 
spectral flow. We will emphasize that the
spectral flow needs to be supplemented with boundary conditions at a cutoff in 
momentum space and from this we
will give a physical picture of the consistent anomaly via anomaly inflow. 
In section 3 will use the Landau level quantization to derive the anomalous 
transport formulas for chiral magnetic and
chiral vortical effects. 
In section 4 will briefly review relativistic hydrodynamics with anomalies and
the fact that the contribution of the (mixed) gravitational anomaly cannot be 
fixed by hydrodynamic arguments alone
due to a mismatch in the number of derivatives in the transport phenomena and 
the anomaly. 
Section 5 will introduce a simple holographic model allowing to make the 
relation between anomalies and
transport coefficients manifest. The derivative mismatch for the gravitational 
anomaly contribution is overcome in holography 
by taking derivatives in the holographic direction.
Section 6 is devoted to the physics of Weyl semi-metals. After a quick 
introduction we will
show how almost all exotic Weyl semimetal phenomenology can be understood from 
anomalous transport theory
as outlined in the previous sections. These include negative magnetoresistivity 
in magnetic fields, in axial magnetic fields,
thermal hall transport and the appearance of edge currents related to Fermi 
arcs. 
Section 7 will then briefly review a recently developed holographic model of 
Weyl semimetal and show that it can be used
to derive a new transport phenomenon related to the gravitational anomaly 
not contained in the ones discussed previously: odd 
viscosity.

\section{Triangle Anomalies}
Let us start with a massless Dirac fermion 
\begin{equation}
 \Psi = \left( \begin{array}{c} \psi_\alpha \\ \bar\phi^{\dot{\alpha}} 
\end{array} \right)\,.
\end{equation}
In a chiral (Weyl-) representation of the $\gamma$-matrices such that
\begin{equation}
 \gamma_5 = \left(  \begin{array}{cc} 1 & 0 \\ 0 & -1\end{array}  \right)\,.
\end{equation}
We define the left- and right-handed spinors via the projector $\mathcal{P}_{\pm 
} = \frac 1 2 \left( 1 \pm \gamma_5 \right)$.
The massless Dirac equation 
\begin{equation}
 i \gamma^\mu  \partial_\mu  \Psi =0\,.
\end{equation}
has two independent $U(1)$ symmetries acting as 
$\psi_\alpha \rightarrow e^{i\varphi_+} \psi_\alpha$ and 
$\bar\phi^{\dot{\alpha}} \rightarrow e^{i\varphi_-} \bar\phi^{\dot{\alpha}}$ 
which
we denote with $U(1)_{L,R}$. The corresponding conserved currents are $J^\mu  
_{L,R} = \bar \Psi \gamma^\mu  \mathcal{P}_\pm \Psi$,
and on the level of classical field theory
\begin{equation}
 \partial_\mu   J^\mu  _{L,R} =0\,.
\end{equation}
For future reference let us also write down the Hamiltonians for left- and 
right-handed fermion in momentum space
\begin{equation}
 \mathcal{H}_\pm = \pm   \vec{p}\vec{\sigma}\,,
\end{equation}
which will be convenient once we discuss Weyl semi-metals.
 \begin{figure}[ht]
\begin{center}
\includegraphics[width=0.8\textwidth]{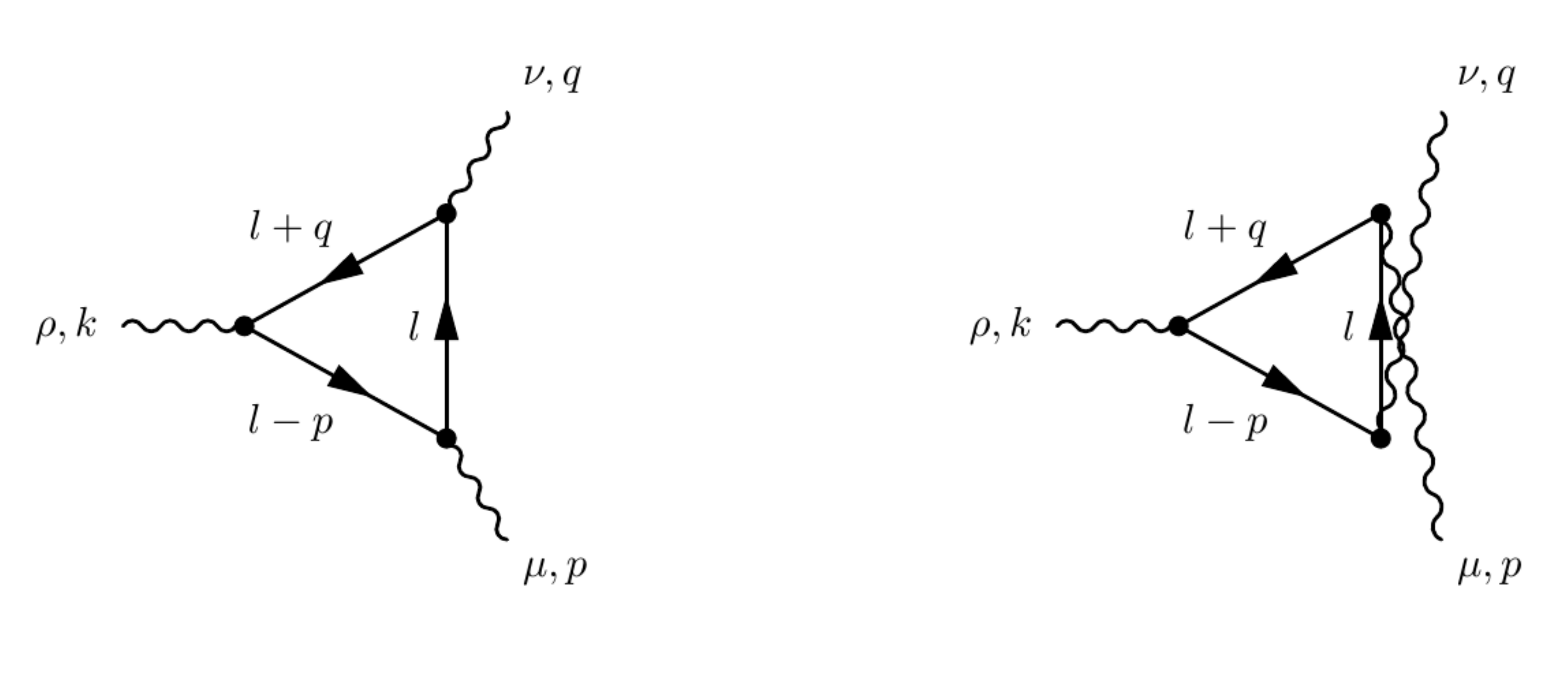}
\caption{Triangle diagrams with three currents at the vertices. While the 
diagrams are linearly divergent the sum is actually finite but undetermined. 
Physical conditions such as Bose symmetry on the external legs (chiral fermions) 
or covariant coupling to the external field (covariant anomaly) or
conservation of the vector current (axial anomaly) have to be imposed in order 
to fix the ambiguity.}
\label{fig:triangle}
\end{center}
\end{figure}
\subsection{Chiral Anomalies}
Let us focus now on a single chiral  fermion. We can define the generating 
functional that allows to compute arbitrary 
n-point functions of the current via gauging. We introduce an external gauge 
field $A_\mu  $ and write the action  as 
\begin{equation}
 S_{\pm } = \int d^4x\, i \bar  \Psi \gamma^\mu  \left (\partial_\mu  - i A_\mu  
\right) \mathcal{P}_{\pm } \Psi \,.
\end{equation}
The quantum effective action (1-particle irreducible) is defined 
as\footnote{More precisely one might use the action
$S_+[A] + S_-[0]$ in the exponent of the path integral to get a well defined 
Dirac operator.}
\begin{equation}
 e^{i \Gamma[A] } = \int D\Psi D\bar\Psi e^{i S_{\pm }}\,.
\end{equation}
Since the action is invariant under the transformation $\delta A_\mu  = 
\partial_\mu  \lambda$ for arbitrary
functions $\lambda(x)$ it follows seemingly that the quantum action is invariant 
and obeys
\begin{equation}
 \int d^4x\, \partial_\mu  \varphi_\pm \frac{\delta \Gamma_\pm [A]}{\delta A_\mu 
 } =0 \,.
\end{equation}
By construction functional variation with respect to the gauge field inserts the 
operator $J^\mu_{\pm }$. Therefore gauge invariance suggests
that $\partial_\mu J^\mu_\pm = 0 $ as an operator equation, i.e. that arbitrary 
correlation functions with one insertion of the divergence of 
the chiral current should vanish. As is well-known this is not true and the 
obstruction of defining such a gauge invariant quantum action
is the chiral anomaly. 

Let us reconsider the anomaly in the elementary triangle diagram of three chiral 
currents.
Applying the usual Feynman rules to the triangle diagrams in figure 
\ref{fig:triangle} we find the three point amplitude ($p+q+k=0$)
\begin{equation}
 i V^{\mu  \nu\rho}(p,q,k)_\pm = \int \frac{d^4l}{(2\pi)^4} \frac{ \mathrm{tr} [ 
(-\slashed l + \slashed p) \gamma^\mu  (-\slashed l) \gamma^\nu(-\slashed l - 
\slashed p) \gamma^{\rho} \mathcal{P}_\pm ]}{(l-p)^2 l^2 (l+q)^2} + 
\left( \mu  \leftrightarrow \nu,  p \leftrightarrow q\right)\,.
\end{equation}
Details of the evaluation of this diagram are discussed in many textbooks such 
as \cite{Srednicki:2007qs} so we will only sketch the most important features. 
First we note that the parity odd part of the projection operator
$\mathcal{P}_\pm $ is relevant. So we replace it with $\frac 1 2 \gamma_5$,
then use  
$\mathrm{tr}[\gamma^\mu \gamma^\nu\gamma^\rho\gamma^\lambda\gamma_5] = 
-4 i \epsilon^{\mu  \nu\rho\lambda}$.
Computing the divergence 
\begin{equation}
 k_\rho V^{\mu  \nu\rho} = 2 \int \frac{d^4l}{(2\pi)^4}  \left[  \frac{ l_\alpha 
(l-p)_\beta}{(l-p)^2 l^2} -   \frac{ l_\alpha (l+q)_\beta}{(l+p)^2 l^2}   
\right] \epsilon^{\alpha\nu\beta\mu  }+ 
\left( \mu  \leftrightarrow \nu,  p \leftrightarrow q\right) =0\,.
\end{equation}
Because of Lorentz symmetry the integrals has to be proportional to either 
$p_\alpha p_\beta$ or $q_\alpha q_\beta$ and these combinations vanish once 
contracted with
the epsilon tensor. On the other handed
\begin{equation}
 p_\mu V^{\mu  \nu\rho} = 2 \int \frac{d^4l}{(2\pi)^4}  \left[  \frac{ l_\alpha 
q_\beta}{(l+q)^2 l^2} -   \frac{ (l-p)_\alpha (p+q)_\beta}{(l-p)^2 (l+q)^2}   
\right] \epsilon^{\alpha\nu\beta\mu  }+ 
\left( \mu  \leftrightarrow \nu,  p \leftrightarrow q\right)\,.
\end{equation}
If the integrals were well defined we could make the substitution $l\rightarrow 
l+p$ in the second integral. The single integral is however linearly divergent 
and has to be defined
properly. Vanishing of $p_\mu   V^{\mu  \nu\rho}$ depends now on the way we have 
labeled the internal loop momentum. Any other choice is just as good. The most 
general choice
is $l \rightarrow l + c(p-q) + d(p+q)$ where $c,d$ are arbitrary real numbers. 
Now the integrals can be evaluated in a Lorentz invariant fashion. All 
divergences cancel but the
final result is undetermined because of the ambiguity in labeling the internal 
loop variable (since the gamma matrix trace gives an epsilon tensor it is only 
the anti-symmetric combination
of external momenta that contributes). One finds
\begin{align}
 p_\mu   V^{\mu  \nu\rho}_{\pm }  & = \pm \frac{-i}{8\pi^2} (1-c) 
\epsilon^{\nu\rho\alpha\beta} q_\alpha k_\beta \,,\\
 q_\nu V^{\mu  \nu\rho}_{\pm }  & =  \pm \frac{-i}{8\pi^2} (1-c) \epsilon^{\mu  
\rho\alpha\beta} k_\alpha p_\beta \,,\\
 k_\rho V^{\mu  \nu\rho}_{\pm }  & =  \pm \frac{-i}{8\pi^2}  2c\, \epsilon^{\mu  
\nu\alpha\beta} q_\alpha p_\beta \,.
\end{align}
Thus the one-loop three point function of three chiral currents is finite but 
undetermined. This poses the question {\em what is the correct value of $c$}? 
Not too surprisingly the
answer to this question is: it depends! It does depend on the physical 
constraints the three point function shall obey. 

First let us go back to the quantum effective action and demand that the three 
point function of currents is 
\begin{equation}\label{eq:Bosesymmetry}
V^{\mu  \nu\rho} = \Gamma_{(3)}^{\mu  \nu\rho} = \frac{\delta^3\Gamma}{\delta 
A_\mu  \delta A_\nu \delta A_\rho}\,.
\end{equation}
Since the order of differentiation does not play any role we must impose Bose 
symmetry on the external legs, all three vertices couple in precisely the same 
way to the gauge field.
This imposes $c=1/3$. If we express the anomaly now in terms of the current and 
the external gauge fields we find
\begin{equation}\label{eq:chiralconsistentanomaly}
\boxed{ \partial_\mu  \mathcal{J}^\mu  _{L,R} = \pm \frac{1}{96\pi^2} 
\epsilon^{\mu  \nu\rho\lambda} F_{\mu  \nu} F_{\rho\lambda} }\,.
\end{equation}
For reasons to be explained shortly this is called the {\em consistent} form of 
the anomaly. 

On the other hand we might be interested to define a quantum operator 
$J^\mu  _\pm $ that has nice properties with respect to gauge transformations. 
More precisely
we would like to think of the current as an object that couples covariantly 
(i.e. without anomaly) to the external gauge fields. This singles out one 
particular vertex and demands
that the divergence on the other two vertices vanishes. The solution for this 
covariant definition of current is $c=1$ and the anomaly is now
 \begin{equation}\label{eq:chiralcovariantanomaly}
\boxed{ \partial_\mu   J^\mu  _{L,R} = \pm \frac{1}{32\pi^2} \epsilon^{\mu  
\nu\rho\lambda} F_{\mu  \nu} F_{\rho\lambda} }\,.
\end{equation}
This looks almost the same as before except for the overall factor of $3$ in the 
anomaly. It is called the {\em covariant} anomaly. 
Since now we have treated the vertices in different ways it is clear that this 
definition of three point amplitude violates the Bose symmetry 
(\ref{eq:Bosesymmetry}). This means
that the covariant current obeying the covariant anomaly equation 
(\ref{eq:chiralcovariantanomaly}) can not be thought of as a functional 
variation of a quantum effective action\footnote{It also appears as the edge 
current in systems where the anomaly
is localized on a co-dimension one boundary and canceled via a higher 
dimensional Chern-Simons term as in quantum Hall systems. See also 
\cite{Jensen:2013kka} for an application of this in the context of anomalous 
transport theory.}.
It might look surprising that we obtained two different answers for the 
divergence of ``the current'' by imposing two different but equally reasonable 
looking conditions. Later when
discussing anomalies and transport we will suggest physical interpretations of 
these different quantum operators, the consistent ($\mathcal{J}^\mu  $) and
the covariant  ($J^\mu  $) currents. 

\subsection{Axial Anomaly}
On the level of classical physics a Dirac fermion is the direct sum of  left- and a 
right-handed chiral fermions, $\Psi_D = \psi_L \oplus \psi_R$. Anomalies pose a 
restriction on
the possibility of defining chiral fermions in the quantum theory. Not too 
surprisingly they also have implications on this direct sum. Let us proceed 
naively
and simply define the quantum theory of a Dirac fermion as the quantum theory of 
a left-handed and a right-handed fermion. We want to keep the external
gauge fields distinguishable, i.e. we introduce left- and right-handed gauge 
fields coupling to the chiral currents independently. We define vector and
axial currents via 
\begin{align}
\mathcal{J}^\mu  &= \mathcal{J}^\mu  _L +  \mathcal{J}^\mu  _R\,,\\
\mathcal{J}_5^\mu  &= \mathcal{J}^\mu  _L -  \mathcal{J}^\mu  _R\,,
\end{align}
 and a basis of vector-like and axial gauge fields 
\begin{align}
A_\mu  = \frac{1}{2} ( A^L_\mu  + A^R_\mu  )\,,\\
 A^5_\mu  = \frac{1}{2} ( A^L_\mu  - A^R_\mu  ).
\end{align}
We can just add and subtract equations (\ref{eq:chiralconsistentanomaly}) to find
\begin{align}\label{eq:anomalyconsistentvector}
 \partial_\mu  \mathcal{J}^\mu  &= \frac{1}{48\pi^2} \epsilon^{\mu  
\nu\rho\lambda} F_{\mu  \nu} F^5_{\rho\lambda} \,, \\
 \partial_\mu  \mathcal{J}_5^\mu  &=  \frac{1}{24\pi^2} \epsilon^{\mu  
\nu\rho\lambda} \left( F_{\mu  \nu} F_{\rho\lambda}+ F^5_{\mu  \nu} 
F^5_{\rho\lambda} \right) \,,
\end{align}
where $F_{\mu  \nu}=\partial_\mu   A_\nu-\partial_\nu A_\mu  $ and $F_{\mu  
\nu}=\partial_\mu   A^5_\nu-\partial_\nu A^5_\mu  $. 
We have chosen to express this in terms of the consistent currents. The result 
is the same up to an overall factor of $3$ for the covariant currents. 
Equation (\ref{eq:anomalyconsistentvector}) looks troublesome: eventually one 
would like the vector current to play the role of the electric current that 
couples
to a dynamical gauge fields. Even without quantizing the gauge fields one should 
expect that the electric current acts as source of Maxwell's equations
\begin{equation}\label{eq:maxwell}
 \mathcal{J}^\mu  = \partial_\nu F^{\mu  \nu}\,.
\end{equation}
This is only consistent if the divergence of the vector current 
vanishes, since $\partial_\mu  \partial_\nu F^{\mu  \nu}=0$. One might say that 
this is still true in
the absence of axial gauge fields and that indeed in nature on a fundamental 
level axial gauge fields do not exists. However one should keep in mind that equation 
(\ref{eq:anomalyconsistentvector})
is just a short form for insertions of the divergence of the vector current in 
correlation functions. Its meaning is that there is a three point function of a 
divergence of a vector current
with an axial current and another vector current that does not vanish. 
Furthermore as we will see later in Weyl semi-metals such axial gauge fields do 
arise quite naturally in the
low energy effective description of their electronics. So we need to solve this 
problem of non-conservation of the vector current. Happily this has been done 
long time ago \cite{Bardeen:1969md}
by noticing that once gauge invariance is lost, nothing prevents us from 
introducing additional (non-gauge invariant) local counterterms to our quantum 
action. 
These are called Bardeen counterterms and they redefine the quantum action as 
follows
\begin{equation}\label{eq:Bardeencts}
 \Gamma[A,A^5] \rightarrow \Gamma[A,A^5] + \int d^4x\, \epsilon^{\mu  
\nu\rho\lambda} A_\mu   A^5_\nu \left( c_1 F_{\rho\lambda} + c_2 
F^5_{\rho\lambda} \right) \,.
\end{equation}
If we now compute the (consistent) currents as variation of the effective action 
with respect to the gauge fields and chose $c_1 = 
\frac{1}{12\pi^2}$ and $c_2=0$ we find 
\begin{subequations}
\begin{empheq}[box=\widefbox]{align}
\partial_\mu  \mathcal{J}^\mu  &= 0 \,,\\ \label{eq:axialanomaly}
  \partial_\mu  \mathcal{J}_5^\mu  &=  \frac{1}{48\pi^2}  \epsilon^{\mu  
\nu\rho\lambda} \left( 3 F_{\mu  \nu} F_{\rho\lambda} + F^5_{\mu  \nu} 
F^5_{\rho\lambda} \right)\,.
\end{empheq}
\end{subequations}
This form of the anomaly is the consistent axial anomaly. The Bardeen 
counterterms guarantee that a conserved  vector current can be defined 
always, independently of the chosen regulator. We have not specified the 
regulator but generically a left-right symmetric regularization would not give a 
conserved vector current.
On the other hand manifestly gauge invariant regulators such as dimensional 
reduction automatically produce the Bardeen counterterms in the 
effective action
and nothing has to be added ``by hand''.  The particular Chern-Simons terms that 
are the Bardeen counterterms exist only if there are at least two independent 
gauge fields. 
That makes the nature of the axial anomaly as a mixed anomaly manifest. The 
precise statement of the axial anomaly is that there is no quantum theory in 
which both the
axial and the vector like currents are conserved at the same time. 

\subsection{Wess-Zumino consistency condition}
So why the anomaly is called {\em consistent}? To understand this we need some 
more formalism. Our object of interest is the quantum effective action 
$\Gamma[A]$
(for simplicity of notation we go back to the case of only one abelian gauge 
field). An anomaly is a non-invariance of the effective action under a gauge 
transformation.
The gauge transformation can be written as a functional differential operator 
$\delta_\lambda = \int d^4x\, \partial_\mu  \lambda \frac{\delta}{\delta A_\mu  
}$ and the anomaly is expressed
as 
\begin{equation}
 \delta_\lambda \Gamma[A] = \mathcal{A}_\lambda\,.
\end{equation}
The right hand side arises because there is no regularization scheme that is 
compatible with the symmetry. It is a remnant of the regularization and remains 
even if we renormalize
and take the regulator to infinity. That makes it intuitively clear that it has 
to be the integral of a local expression in the field $A_\mu$. On the other hand 
the quantum effective action
arises by integrating out massless (chiral) fermions and is essentially a 
non-local expression. We further observe that the gauge transformations have to 
obey the gauge algebra
which in our simple example means that two gauge transformations with different 
gauge parameters have to commute.
\begin{equation}
 [ \delta_\lambda, \delta_\sigma ] = 0\,.
\end{equation}
It follows now that the anomaly has to fulfill
\begin{equation}
 \delta_\lambda \mathcal{A}_\sigma - \delta_\sigma \mathcal{A}_\lambda =0\,.
\end{equation}
This is the Wess-Zumino consistency condition. A more geometric formulation can 
be given with one more piece of formalism. Let us promote the gauge parameters
to a Grassmann valued field $\lambda(x) \rightarrow c(x)$ called the 
``ghost''\footnote{In the path integral quantization of non-abelian gauge 
theories this is the Fadeev-Popov ghost
that arises in defining the measure.}. It is useful to have the analogy of the 
exterior derivative $d$ and the formalism of differential forms in mind,
e.g. the field strength of a gauge field (1-form) $A=A_\mu   dx^\mu  $ is 
defined as the exterior derivative $F=\frac{1}{2} F_{\mu  \nu} dx^\mu   dx^\nu = 
dA$. 
In an analogous way let us introduce an exterior derivative on field space 
\begin{equation}
s=\int d^4x\,\partial_\mu   c \frac{\delta}{\delta A_\mu  }\,.
\end{equation}
Which is nothing but a gauge transformation with the Grassmann valued ghost 
field $c$ as gauge parameter. 
It is called the BRST operator\footnote{See \cite{Dragon:2012au} for
a recent review.}. 
As one can check easily it is nilpotent $s^2=0$. The anomaly can
now be written as 
\begin{equation}
 s \Gamma[A] = \mathcal{A}\,.
\end{equation}
We can think of the anomaly as a one-from on field space (a local integrated 
polynomial of the field $A_\mu  $, one ghost field and a finite number of 
derivatives, i.e. having ghostnumber one).
The Wess-Zumino consistency condition is the fact that the anomaly is a closed 
one-from with respect to the BRST operator 
\begin{equation}
 s \mathcal{A} =0\,~~~,~~~ \mathcal{A}\neq s \Gamma_{c.t.}[A]\,.
\end{equation}
Here we have included the condition that it should not be possible to write the 
anomaly as the BRST variation of an integral of local term of ghostnumber zero. If that 
were the case we could just
add $\Gamma_{c.t.}$ as counterterm of the effective action and get a new 
redefined and BRST (gauge-) invariant quantum action. This maps the anomaly to a 
cohomology problem:
the consistent anomaly is a non-trivial element of the BRST cohomology on the 
space of local integrated monomials in the fields at ghostnumber one. 
Finally let us note that all this formalism can be extended in full generality 
to non-abelian gauge algebras.  

\subsection{Covariant Anomaly}
OK, so now we know that the consistent anomaly is a solution to the consistency 
condition. But what is the {\em covariant} anomaly?
Well, we observe that the consistent current defined as the functional 
derivative of the quantum action is not a gauge invariant operator if there is 
an anomaly.
Using $[s, \delta/\delta A_\mu  ] =0$ we find
\begin{equation}\label{eq:gaugevariantioncurrent}
 s \mathcal{J}^\mu  = \frac{\delta}{\delta A_\mu  } \mathcal{A}= -\frac{\pm 
1}{24\pi^2}\epsilon^{\mu  \nu\rho\lambda} \partial_\nu c F_{\rho\lambda},.
\end{equation}
where in in the expression on the right we have specialized to one chiral 
fermion again. 
We already know that in the triangle diagram we can put all the anomaly 
into a single vertex. So it must be possible to define a current with 
covariant couplings to the external legs. In particular we should demand from this 
quantum operator $s J^\mu  =0$ even in the presence of an anomaly. 

From (\ref{eq:gaugevariantioncurrent}) it is easy to see that by adding a 
Chern-Simons current to the consistent current
we can define 
\begin{equation}\label{eq:covchiralcurrent}
 J^\mu  _{L,R} = \mathcal{J}^\mu  _{L,R} + \frac{\pm 1}{24\pi^2} \epsilon^{\mu  
\nu\rho\lambda} A_\nu F_{\rho\lambda},.
\end{equation}
Adding the Chern-Simons current to the consistent 
current we can construct the covariant current. The defining characteristics of 
this current are
that it is invariant under all the gauge transformation, even the anomalous ones, 
and that it can not be obtained from variation of an action with local 
counterterms. We emphasized
this already in the analysis of the triangle diagram but now we can also see it 
from he Chern-Simons current in (\ref{eq:covchiralcurrent}). 
We can compute the anomaly in the covariant current (it is a covariant 
object under the anomalous gauge transformations but it does have an anomaly by 
itself)
\begin{equation}
 \partial_\mu J^\mu  _{L,R} = \frac{\pm 1}{32\pi^2}\epsilon^{\mu\nu\rho\lambda}F_{\mu  \nu} F_{\rho\lambda}\,.
\end{equation}
As expected this is $1/3$ of the consistent anomaly.\footnote{We could also define 
a conserved current by adding the Chern-Simons current with an appropriate coefficient.
Such a current is then neither consistent (variation of an effective action) nor
gauge invariant.}

We can go through the same exercise in the case of the axial anomaly 
and construct the covariant vector and axial currents. With our canonical 
choice 
of taking the vector current explicitly conserved we have

 \begin{subequations}
 \begin{empheq}[box=\widefbox]{align}\label{eq:covcurrent}
 J^\mu  &= \mathcal{J}^\mu  + \frac{1}{4\pi^2} \epsilon^{\mu  \nu\rho\lambda} 
A^5_\nu F_{\rho\lambda}\,,\\ \label{eq:covaxialcurrent}
 J_5^\mu  &= \mathcal{J}_5^\mu  + \frac{1}{12\pi^2} \epsilon^{\mu  
\nu\rho\lambda} A^5_\nu F^5_{\rho\lambda}\,.
 \end{empheq}
 \end{subequations}
Note that only the axial gauge potential enters these expressions. This is a 
reflection of the fact that we have chosen to put all the anomaly into the
axial current. 
For future reference let us also write down the covariant vector and axial 
anomaly
 \begin{subequations}
 \begin{empheq}[box=\fbox]{align}\label{eq:covanomalies}
 \partial_\mu   J^\mu  &= \frac{1}{8\pi^2}  \epsilon^{\mu  \nu\rho\lambda}  
F_{\mu  \nu} F^5_{\rho\lambda} &= \frac{1}{2\pi^2} \left( \vec E .\vec B_5 + 
\vec{E}_5. \vec{B}\right) \,,\\
 \partial_\mu   J_5^\mu  &= \frac{1}{16\pi^2}  \epsilon^{\mu  \nu\rho\lambda} 
\left( F_{\mu  \nu} F_{\rho\lambda}+F^5_{\mu  \nu} F^5_{\rho\lambda}\right) &= 
\frac{1}{2\pi^2} \left( \vec E .\vec B + \vec{E}_5. \vec{B}_5\right) \,.
 \end{empheq}
 \end{subequations}
The covariant anomaly looks completely vector-axial symmetric 
once we express it in therms of electric and magnetic fields. 
Generally there always exists Chern-Simons currents that can be added to the 
consistent currents making the resulting covariant current a covariant object under 
all gauge transformations.
They are also known as Bardeen-Zumino polynomials (not to
be confused with the Bardeen counterterms). The theory of covariant and 
consistent anomalies goes back to
\cite{Bardeen:1984pm}. 

\subsection{Gravitational Anomaly}
There is one more anomaly that appears in the triangle diagram of one chiral 
current and two energy-momentum tensors. This is the gravitational contribution
to the chiral anomaly (also 
mixed gauge-gravitational)
anomaly \cite{Eguchi:1976db, Delbourgo:1972xb, AlvarezGaume:1983ig}. It is by 
nature a mixed anomaly and therefore one can always use Bardeen counterterms to 
shift the anomaly between the involved symmetries. Note that the
Bardeen counterterms have the form 
connection$\wedge$connection$\wedge$field-strength. In the case of gravity the 
connection (=gauge field) is the Levi-Civita connection
and the field strength is the Riemann tensor. On a fundamental level, gravity 
is always gauged in nature and that implies that there should not be any anomaly 
in the diffeomorphism
symmetry. So it is customary to shift the anomaly completely into the chiral 
current in which case it takes the form
\begin{equation}
 \nabla_\mu  \mathcal{J}^\mu  = \frac{\pm 1}{768\pi^2} \epsilon^{\mu  
\nu\rho\lambda} R^\alpha\,_{\beta\mu  \nu} R^\beta\,_{\alpha\rho\lambda} \,.
\end{equation}
for a single chiral fermion. Again one can find a covariant form of this anomaly 
applying the principles outlined in the previous subsection. 
While this looks a rather straightforward application of the principle that 
anomalies are contractions of field strength tensors with
the epsilon tensor there is at least one clear difference: the usual chiral and 
axial anomalies are expressions involving two derivatives whereas the
gravitational anomaly involves four derivatives. 

\subsection{Anomaly coefficients}
Anomalies are subject to non-renormalization theorems. The 
anomaly coefficient is exact at the one loop level. So one can infer the 
presence
of an anomaly by analyzing the triangle diagram with generic currents at the 
vertices. Let us assume a generic symmetry group generated by matrices $T_a$
such that $[T_a, T_b] = i f_{abc} T_c$. Chiral anomalies are present if the 
anomaly coefficient
\begin{equation}
 d_{abc} = \frac 1 2 \sum_l  \mathrm{tr}\left( \{T^l_a,T^l_b\} T^l_c \right)  -  
\ \frac 1 2 \sum_r  \mathrm{tr}\left( \{T^r_a,T^r_b\} T^r_c\right) \,,
\end{equation}
does not vanish. Here the sums run over the species of left- and right-handed fermions and 
$T^{l.r}$ are the representations of left- and right-handed fermions and the 
curly bracket
is the anti-commutator.
In the case when all symmetries are abelian this boils down to sums over triple 
products of charges
 \begin{equation}
 d_{abc} =  \sum_l (q_a^l q_b^l q_c^l ) -  \sum_r (q_a^r q_b^r q_c^r )\,.
\end{equation}
 A mixed gravitational anomaly is present if 
\begin{equation}
 b_a = \sum_l q_a^l - \sum_r q_a^r
\end{equation}
is different from zero. We call $d_{abc}$ and $b_a$ the chiral and gravitational 
anomaly coefficients. Note that $d_{abc}$ is completely symmetric. By means
of adding Bardeen counterterms one can cancel some of the consistent anomalies. This is
precisely the case of the axial anomaly where $d_{AVV}=d_{VAV}=d_{VVA}\neq0$ but the
consistent vector current is conserved. 

\section{Landau Levels and anomalies}
\begin{figure}[ht]
\begin{center}
\includegraphics[width=0.6\textwidth]{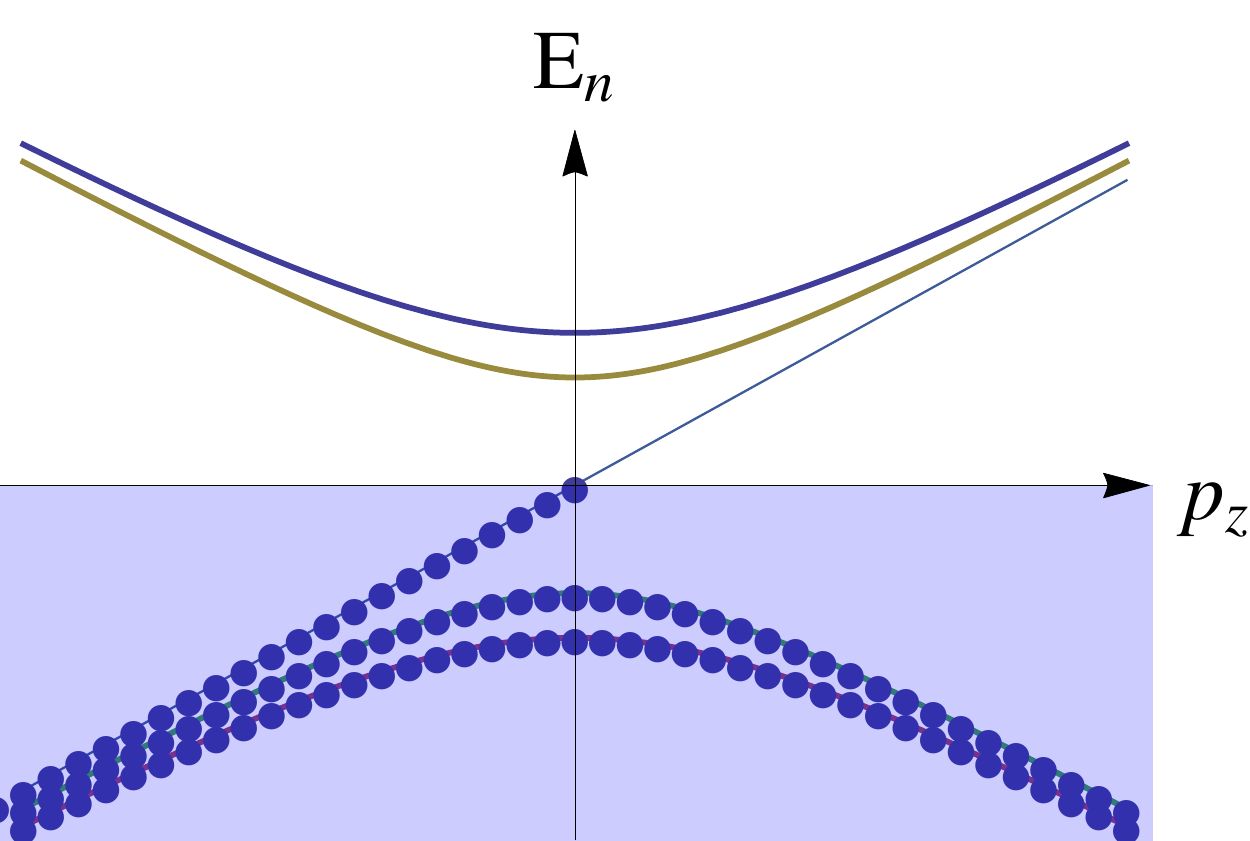}
\caption{Landau level spectrum of s single Weyl fermion. The higher Landau 
levels are spin degenerate and gapped. The lowest Landau level is chiral fermion 
whose motion is
restricted along the magnetic field. The Dirac sea is comprised of all the 
states of negative energy which includes all lowest Landau level states of 
negative momentum.
The spin is polarized along the magnetic field.}
\label{fig:LLspectrum}
\end{center}
\end{figure}

\begin{figure}[ht]
\begin{center}
\includegraphics[width=0.6\textwidth]{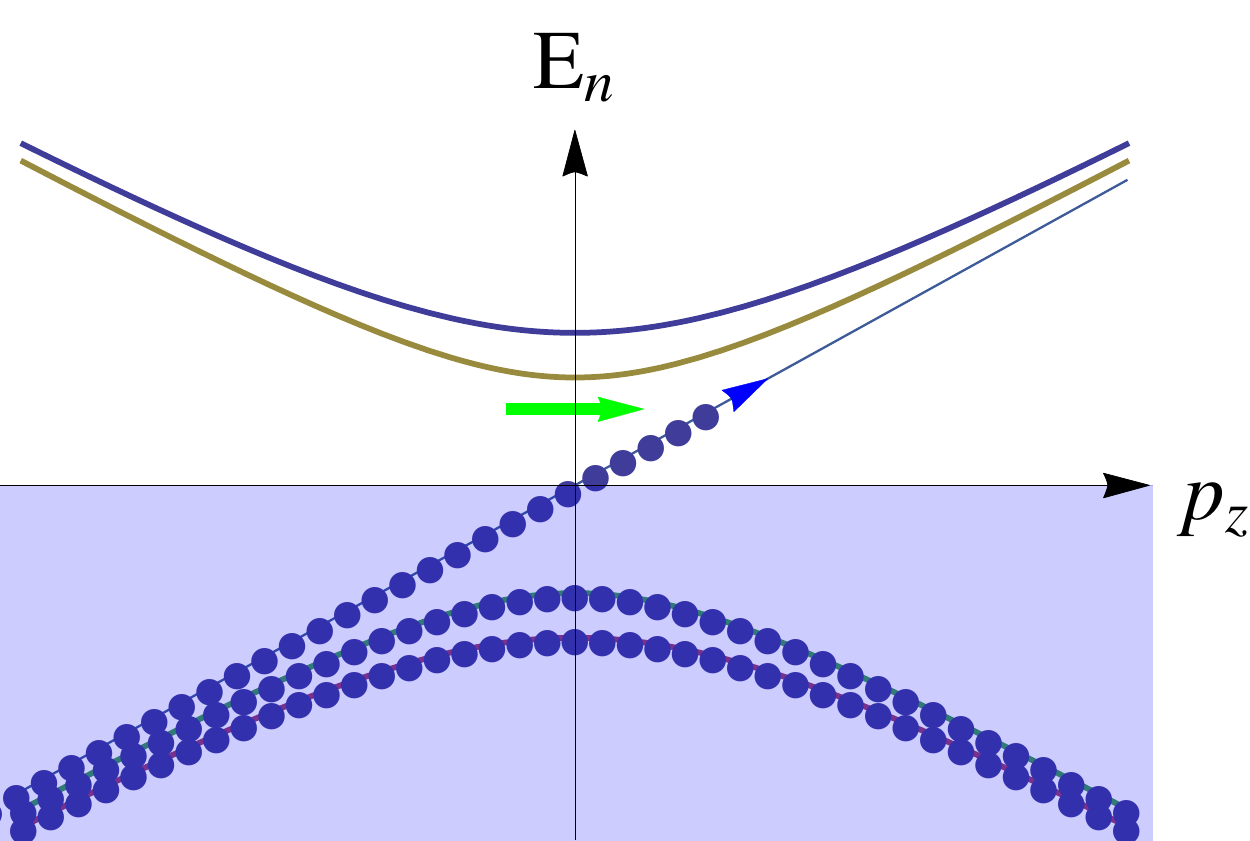}
\caption{Spectral flow picture of the chiral anomaly. 
 In parallel electric and magnetic fields the states of the lowest Landau level 
are pushed across the normal ordered
vacuum. The direction of the spectral flow is indicated by the blue arrow. The 
electric field is indicated by the green arrow.
From the quantum field theoretical perspective particles are created out of the 
vacuum. Since there is an infinite supply of states in the Dirac sea it can not
be depleted.}
\label{fig:chiralanomalyspectralflow}
\end{center}
\end{figure}

We now study chiral fermions in a magnetic field\footnote{A recent review on quantum field
theory in magnetic field backgrounds is \cite{Miransky:2015ava}}. The Weyl equation is
\begin{equation}
 i \slashed{D} \Psi =0\,.
\end{equation}
and the covariant derivative is $D_\mu  = \partial_\mu  - i A_\mu  $ (absorbing 
the electric charge into the definition of the gauge field).
The magnetic field is taken to point in the $z$-direction and the gauge field is 
chosen $A_y= B x$. 
The Weyl equation is 
\begin{equation}
 i (\partial_t - \vec\sigma \vec D ) \psi =0\,.
\end{equation}
We now use the fact that the differential equation depends explicitly only on 
$x$ and not on the other coordinates so we can use the ansatz
$\psi = e^{-i(\omega t-p_y y - p_z z)}\tilde\psi(x)$ to find the matrix equation
\begin{equation}
 \left( 
\begin{array}{cc}
 \omega-p_z & i(\partial_x - Bx + p_y) \\
- i(\partial_x + Bx - p_y) & \omega +p_z
 \end{array}
\right) .
\left(
\begin{array}{c}
 \psi_+ \\  \psi_-
\end{array}
\right) =0\,.
\end{equation}
For $\omega=p_z$ there is a simple solution $\psi_+ \propto \exp[ 
\frac{-(B x-p_y)^2}{2B}]$ and $\psi_-=0$ whereas the corresponding solution
with $\omega=-p_z$  and $\psi_+=0$ is non-normalizable. This is 
the Lowest Landau 
with $s=+1$, where $s$ is the eigenvalue of the spinor wave 
function 
of $\vec \sigma . \vec{B}/|B|$.  For a chiral fermion of opposite chirality one 
finds
that the normalizable solution has $\omega=-p_z$ and $s=+1$.
The lowest Landau level $n=0$ disperses linearly  
\begin{equation}
\omega_{p_z,0} = \pm p_z
\end{equation}
with the sign determined by chirality.
The rest of the 
spectrum arranges into Landau levels of positive and negative energies given by 
\begin{equation}
 \omega_{p_z,n}  = \pm \sqrt{p_z^2 + 2 B n}\,,
\end{equation}
with $n = 1,2,3,\dots$. 
The Weyl equation in a magnetic field can be separated into a plane wave and a 
harmonic
oscillator corresponding the the degrees of freedom along and transverse to the 
magnetic field. 
Accordingly the momentum along the magnetic field is still a good quantum number 
but the momenta in the plane transverse the the magnetic field
are replaced by just the harmonic oscillator quantum number $n$. In our gauge 
choice the momentum in $y$ direction parametrizes the degeneracy
of the Landau levels of $\frac{B}{2\pi}$ states per unit area.  
The spectrum is sketched in figure \ref{fig:LLspectrum}.
Let us apply now the standard argument that allows to derive the chiral anomaly 
from spectral flow \cite{Nielsen:1983rb}. In addition to the magnetic field we 
switch on a parallel electric field
$E_z$. This field will pump momentum into the system according to Newtons law 
$\dot{p}_z = E_z$.
In the Dirac sea of the higher Landau levels all states are occupied and
a fermion has no available state to move to\footnote{In condensed matter
physics it is known that fully occupied bands do not produce an electric
current (if not for topological reasons).}.
For the lowest Landau level 
there is something more going on. We assume, as usual in quantum 
field theory, that
the infinite Dirac sea of negative energy states is subtracted via a normal 
ordering prescription. The electric field 
pumps momentum
into the system and shifts the states in the Dirac sea of the of the lowest 
Landau level to positive momentum! Occupied states just below the normal ordered 
vacuum are shifted into
empty states just above the vacuum. This is particle creation out of the 
vacuum. We can also compute the rate of particle creation. The density of 
states for a one dimensional
chiral fermion (such as the fermions in the lowest Landau level) is $dn = 
dp/(2\pi)$ and their degeneracy is $B/(2\pi)$. If we combine this with the 
Lorentz force we find
\begin{equation}
 \frac{dn}{dt} = \frac{\vec E. \vec B}{4\pi^2}\,.
\end{equation}
But this is just a Lorentz non-covariant version of the anomaly equation $\partial_\mu   
J^\mu  = \frac{1}{32} \epsilon^{\mu  \nu\rho\lambda} F_{\mu  \nu} 
F_{\rho\lambda}$!. 
So now the anomaly has been recovered from rather elementary quantum  mechanics 
of a singe Weyl fermion without any fancy quantum field theory. Why was this so
easy and where does quantum field theory hide? It hides in two aspects: first we 
assumed a notion of normal ordered vacuum which is our trick to subtract the
infinite Dirac sea. The anomaly is then the incompatibility between the spectral 
flow and our normal ordering prescription.  There is another aspect to 
it: where do all
the fermions come from? Well in this picture we do not really have to ask this 
question since the Dirac sea is infinite and any finite amount of states that we 
pull out of the
vacuum will not be able to deplete the infinite supply of states in the Dirac 
sea\footnote{Sometimes this is compared to the Hilbert Hotel with an infinite 
number of rooms. Any
new arrival can be accommodated by simply asking the occupants in room number $n$ 
to switch to room number $n+1$ leaving the room number $1$ available.} In 
quantum field
theory there is another ingredient that we will have to take into 
account eventually: a cutoff has to be introduced at intermediate stages 
of the calculations. This will
turn out to be an essential ingredient to the proper spectral flow picture of 
the axial anomaly. For the moment we want to point out that the 
prefactor of the anomaly 
obtained via the spectral flow argument is the one of the {\em covariant }
anomaly. In hindsight this is not surprising, since we assumed that our chiral 
fermions couple 
covariantly to the external fields (i.e. we assumed the usual form  the  Lorentz 
force.)  The spectral flow picture of the chiral anomaly is sketched in figure 
\ref{fig:chiralanomalyspectralflow}.

\begin{figure}[ht]
\begin{center}
\includegraphics[width=0.6\textwidth]{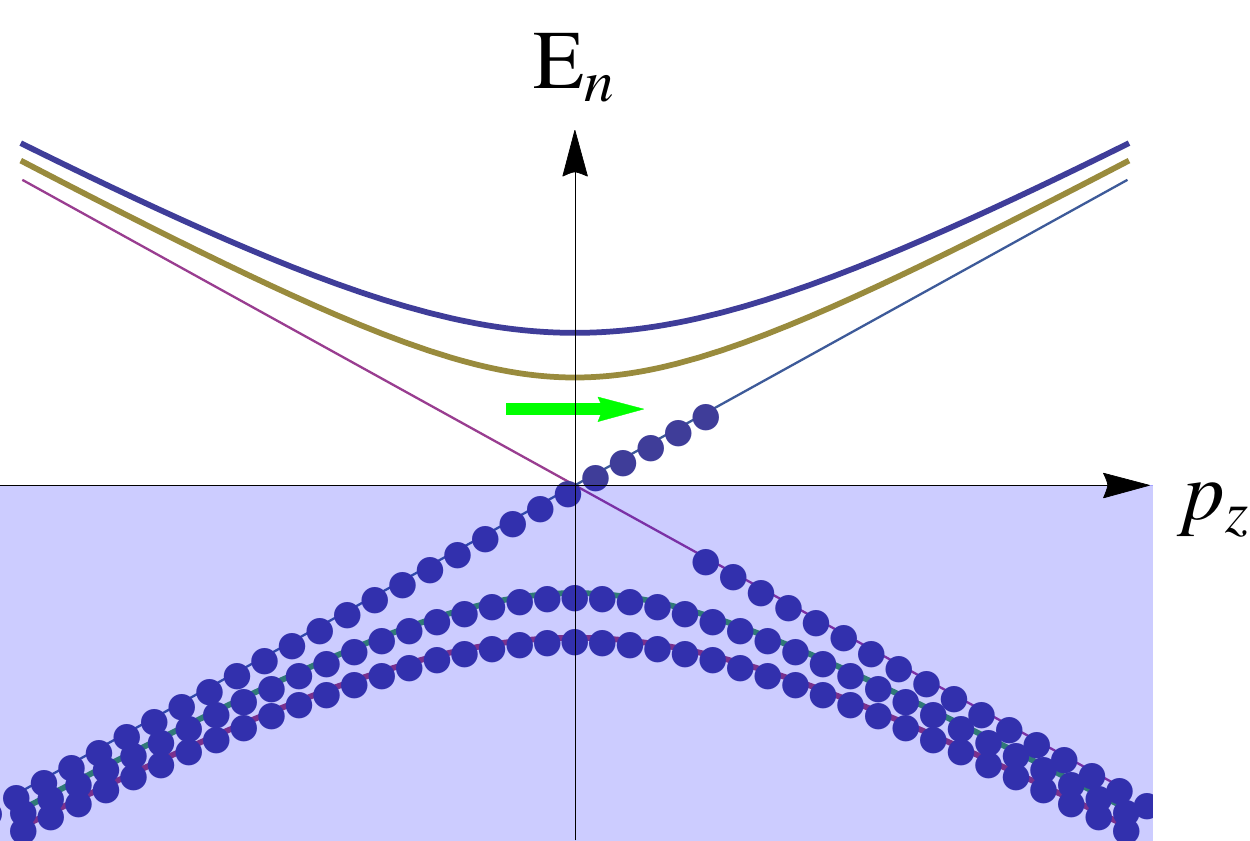}
\caption{Spectral flow picture of the axial anomaly. 
Chiral fermions of both chiralities are present. The particle creation of 
right-handed particles is counterbalanced by annihilation of left-handed 
particles. The total
number of particles does not change but an imbalance in the number of 
right-handed and left-handed fermions is pumped into the system.}
\label{fig:axialanomalyspectralflow}
\end{center}
\end{figure}

Let us now combine left- and right-handed Weyl fermions to get an idea of the 
spectral flow of the axial anomaly. States are created and annihilated
since for each right-handed fermion pulled out of the vacuum 
there is a left-handed particle that is pushed further down into the 
Dirac sea.
The total anomaly is
\begin{equation}
 \frac{d n}{ dt} = \frac{d (n_L - n_R) }{dt} = \frac{1}{2\pi^2} 
\vec{E}.\vec{B}\,.
\end{equation}
For this particular case we can actually not decide if this is the covariant or
the consistent anomaly (\ref{eq:anomalyconsistentvector}) with conserved vector 
current. Both anomalies have the same coefficient, which is not surprising since 
the
vector current actually is anomaly free and trivially a covariant object under 
(vector-like) gauge transformations. The spectral flow picture for the axial 
anomaly is
depicted in figure \ref{fig:axialanomalyspectralflow}.

\begin{figure}[ht]
\begin{center}
\includegraphics[width=1.0\textwidth]{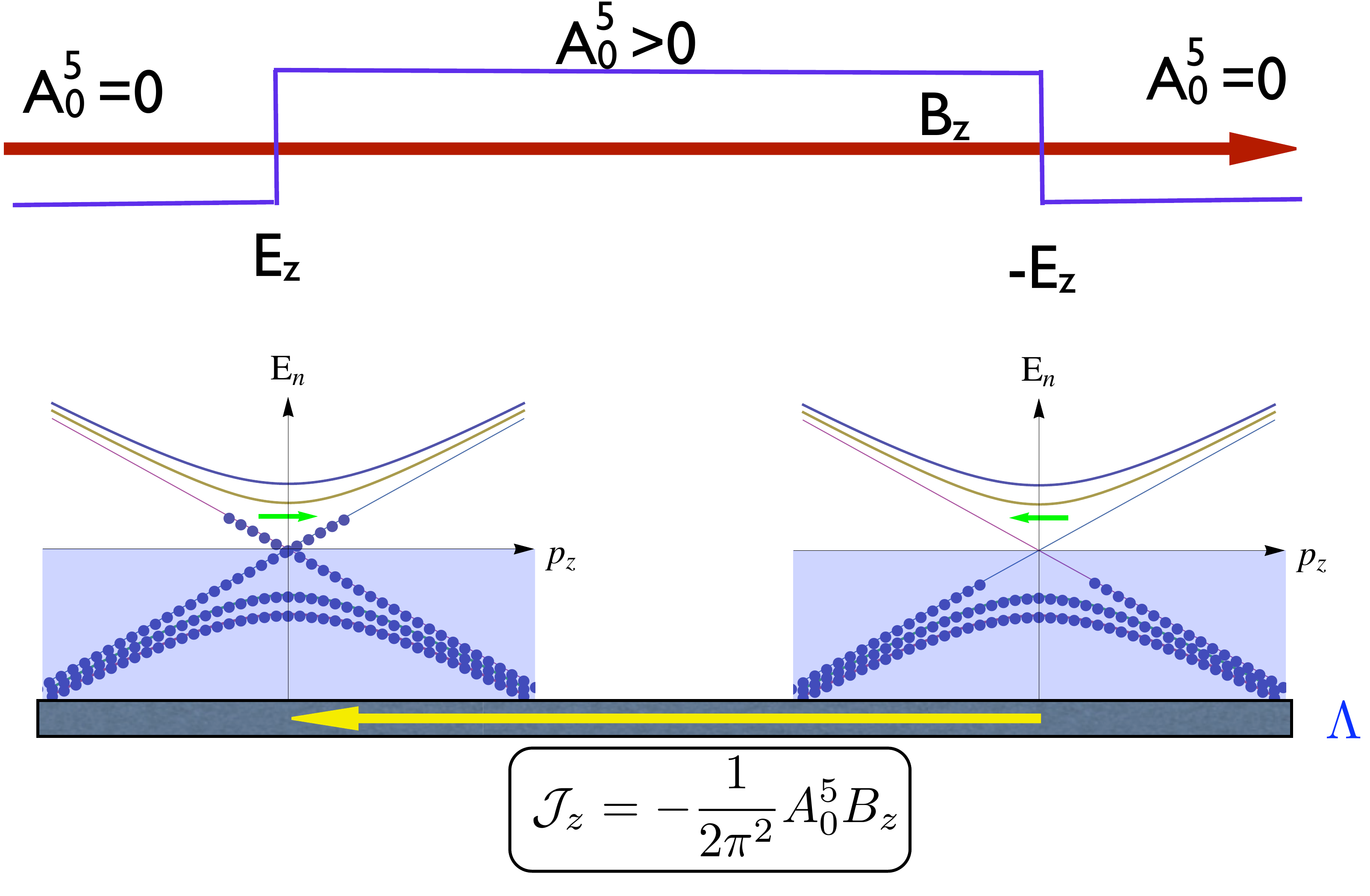}
\caption{Spectral flow picture of the consistent current.  We assume the 
boundary condition $A_0^5 =0$ at infinity and take constant but non-zero value in a 
finite slab
$|z|<L$. At $x=\pm $ strong localized axial electric fields a present. In 
addition we assume a constant (vector-like) magnetic field $B_z$. From the 
perspective of the
covariant anomaly this activates the anomaly in the (covariant) vector current 
at $x=\pm L$ where charges of equal amounts are either created of destroyed. 
The 
boundary condition on the axial  vector field $A_\mu  ^5$ guarantees that no net 
charge is created. The spectral flow of the fermions necessarily hits 
the
cutoff $\Lambda$ in the regulated quantum field theory and is subject to 
boundary conditions there. Charge preserving boundary conditions can be given 
which create
a current whose quantum field theory implementation is the Bardeen-Zumino 
polynomial. This Chern-Simons current generated at the cutoff guarantees local 
charge conservation!}
\label{fig:consistentcurrentflow}
\end{center}
\end{figure}

Now let us see how we can understand the spectral flow when we switch on parallel 
axial electric field and usual (vector-like) magnetic field. The spectral 
flow picture tells
us that both, left-handed and right-handed particles are created out of the 
vacuum. This indeed looks problematic since it would mean that vector-like 
charge (which eventually
we want to identify with the electric charge) is not conserved. To understand 
better what is going on we need to keep in mind that the axial field 
$A_\mu^5$ is of different
nature than the true gauge field $A_\mu$ because of the anomaly. Whereas the 
proper gauge field is not an observable itself $A_\mu^5$ is in principle 
observable. A pure gauge $A^5_\mu= \partial_\mu \lambda_5$ does not decouple
from the theory, rather it couples to the anomaly. 
It is therefore natural to impose the boundary condition $A^5_\mu=0$ at 
infinity, $A_\mu^5$ can be different from zero only in  a 
compact domain.
Let us imagine $A_0^5\neq 0$ but constant in a slab $|z|<L$. Then at 
$x=\pm L$ there is a strong gradient of $A_0^5$ which is nothing but an axial 
electric field $\vec{E}_5$.
If a parallel (vector-like) magnetic field is present the covariant anomaly 
(\ref{eq:covanomalies}) gets excited and creates charge out of the vacuum at 
$z=-L$ but destroys charge
in equal amounts at $z=+L$. So that's already good since globally no net charge 
is created. But it is not yet good enough! Charge conservation is a local 
equation and tells us
that charge can only leave or enter a bounded region via inflow or outflow of 
current. So even with the boundary condition $A_0^5|_\infty =0$ we  still 
violate local charge conservation. This is 
where the Bardeen-Zumino polynomial comes to rescue! The relation between the 
covariant and
conserved consistent current (\ref{eq:covcurrent}) tells us that in a region 
with non-vanishing axial vector $A_\mu^5$ and simultaneous presence of 
vector-like field-strength
a Chern-Simons current is created. This current is precisely such that it 
provides the inflow guaranteeing local charge conservation.
The spectral flow picture of this consistent picture of the ``anomaly'' is 
depicted in figure \ref{fig:consistentcurrentflow}.
The fermions following 
the spectral flow do not really just move out to arbitrary high momentum. In 
quantum field theory we
always have to include a finite cutoff $\Lambda$ at intermediate stages of 
our calculations. So with the regulator in place the fermions subject to 
spectral flow hit the cutoff
and we need to tell them what they shall do at the cutoff by imposing boundary 
conditions in momentum space. Of course the correct boundary condition for a 
conserved electric
current is that as soon as the fermions hit the cutoff they generate a current 
in space transporting precisely the right amount of charge from the region of 
negative axial electric
field to the region of positive axial electric field. This inflow 
guarantees local charge conservation. Of course we could also 
impose some sort of leaky
boundary conditions in which we allow some of the charge to vanish into 
to the region beyond the cutoff. On the level of the effective action $\Gamma$ 
the boundary
conditions at the cutoff are parametrized by the Bardeen counterterms 
(\ref{eq:Bardeencts}). 
This is basically the same mechanism that is behind the anomaly inflow mechanism 
of \cite{Callan:1984sa}. The only difference being that now the bulk is not 
gapped\footnote{In view of applications in condensed matter we might propose 
the maps {\em Callan-Harvey}$\leftrightarrow${\em topological insulators} vs. 
{\em Bardeen-Zumino}$\leftrightarrow${\em topological metals}.}. 

The Chern-Simons current generated to guarantee local charge 
conservation is $\mathcal{J}_z = -\frac{A_0^5}{2\pi^2} B_z$. This looks 
precisely like the celebrated 
chiral magnetic effect, i.e. the generation of a current via magnetic field. It 
is only one part of the CME. So far we have assumed that our 
system was in a
vacuum state with respect to the normal ordered vacuum, i.e. no states above 
the vacuum are occupied. In the region where $A_0^5\neq 0$  the vacuum energies 
are shifted: the left-handed vacuum is shifted down by 
$A_0^5$ and the right-handed vacuum is shifted up by $A_0^5$.  
In the next section we will study another source of magnetic (and rotation) 
induced currents built up by the occupied states above the normal 
ordered vacuum.

\section{Transport from Landau Levels}
For a fermion the vacuum is the state with vanishing Fermi surface.
Let us now go beyond this restriction and study what happens when there is a 
non-vanishing Fermi surface and once we're at it we also introduce finite 
temperature (so 
strictly speaking there is no sharp Fermi surface but it is smoothed by the 
temperature). To start we study what happens if we have a single chiral fermion in 
a magnetic field
with chemical potential $\mu_{L,R}$ at temperature $T$. Of course the charge and 
energy is non-vanishing and this is described by the free energy density\footnote{For 
completely free fermions we also could introduce independent left- and right-handed 
temperatures. If we understand the the expressions as lowest order in a 
perturbative
expansion of an interacting theory only a common temperature can be defined. 
Local interactions can preserve separate left- and right-handed
$U(1)$ symmetries so different chemical potentials can be defined in the 
interacting theory. The anomaly spoils this but only in the simultaneous
presence of electric and magnetic fields.}
\begin{equation}\label{eq:fwithoutB}
 F_{L,R} = -\frac{1}{24\pi^2} \left( \mu  _{L,R}^2 + 2 \pi^2 \mu  _{L,R}^2 T^2 + 
\frac{7}{15} \pi^4 T^4\right)\,.
\end{equation}
if the magnetic field is much smaller than Temperature and chemical potentials 
and by 
\begin{equation}\label{eq:fwithB}
 F_{L.R} = - B \left( \frac{T^2}{24} + \frac{\mu  _{L,R}^2}{8\pi^2} \right) + 
O\left( e^{-\frac{\sqrt{B}-\mu  _{L,R}}{T} } \right)\,,
\end{equation}
for large magnetic field. The last expression is valid as long as the magnetic 
field induces a gap in the higher Landau levels $\sqrt{2 B} \gg (T,\mu  _{L,R} 
)$. In that case
the only states that contribute are the states in the lowest Landau level and 
these are effective $1+1$ dimensional chiral fermions. The contribution from the 
higher Landau levels
in this regime is exponentially suppressed.

Let us compute the current in a magnetic field. First we start with the 
higher Landau levels. Note that the current is simply the integral over the 
velocity of of the
particles weighted by the Fermi-Dirac distribution 
\begin{equation}
 J_{HLL} = \int_{-\infty}^\infty \frac{dp}{2\pi} \frac{\partial E_n}{\partial p} 
\left[ \frac{1}{1+ e^{\frac{E_n-\mu  }{T}} } - \frac{1}{1+ e^{\frac{E_n+\mu  
}{T}} }\right] =0\,.
\end{equation}
The velocity in the higher Landau Levels is the group velocity $\partial 
E_n/\partial p$, both particles and anti-particles contribute, the latter 
however with opposite sign.
The integrand is an odd function of $p_z$. Since the probability of finding a 
particle with infinite momentum
is exponentially suppressed the integral is well defined and evaluates to zero.
There is no current generated in the higher Landau levels.

The lowest Landau level is different. The integration region is only the 
positive half line and the velocity is simply $\pm 1$ depending on chirality. So 
we find
\begin{equation}
 J_{LLL} = \pm \int_0^\infty \frac{dp}{2\pi}  \left[ \frac{1}{1+ e^{\frac{p-\mu_{L,R}  
}{T}} } - \frac{1}{1+ e^{\frac{p+\mu_{L,R}}{T}} }\right]  = \pm \frac{\mu_{L,R}}{2\pi} \,.
\end{equation}
Now we remember that these states come with a multiplicity of $B/(2\pi)$ per unit 
area and this give the current density 
\begin{equation}\label{eq:cmechiral}
 \vec{J} = \pm \frac{\mu_{L,R}}{4\pi^2} \vec{B} \,.
\end{equation}
Finally here it is, the celebrated expression for the chiral magnetic effect:
The generation of a current in a magnetic field. Many things are remarkable 
about this formula. First
we have computed it at finite temperature but the result is completely 
independent of it. Only the lowest Landau level contributes, but contrary to the 
expression for the free
energy  (\ref{eq:fwithoutB}) this is not an approximation valid for large field 
strength but it is an exact result! 

Let us go a step further, let us also compute the energy current. Since we are 
dealing with a relativistic theory we can equally well call it the momentum 
density along the magnetic
field. The calculation is very similar, only the lowest Landau level contributes 
and we find
\begin{equation}
 J_{\epsilon,LLL} =  \pm \int_0^\infty \frac{dp}{2\pi}  \left[ \frac{p}{1+ 
e^{\frac{p-\mu  }{T}} }  + \frac{p}{1+ e^{\frac{p+\mu  }{T}} }\right]  = \pm 
\left( \frac{\mu  _{L,R}^2}{4\pi} + \frac{\pi T^2}{12}\right) \,.
\end{equation}
giving
\begin{equation}\label{eq:cmeenergy}
 \vec{J}_\epsilon = \pm \left( \frac{\mu  _{R,L}^2}{8\pi^2} + \frac{T^2}{24} 
\right) \vec B \,.
\end{equation}
This is the chiral magnetic effect in the energy current. 
Now the temperature contributes but only in a very simple polynomial way!

It is a common lore in many body physics that rotation has many similarities to 
magnetic fields. For example the Coriolis force $\vec F =2 m \vec v \times\vec 
\omega$
is similar to the Lorentz force $\vec F = \vec v \times  \vec  q B$ if we identify $2 
m\vec \omega \sim q\vec  B$\footnote{Of course rotation also gives rise to a 
centrifugal force. 
It is quadratic in the angular velocity $\omega = \vec\nabla \times\vec v$. 
We might try to ignore it on the grounds that it is higher order in 
derivatives.} 
For relativistic, massless fermions we should replace the rest mass with the 
energy. Now we can calculate the current due to rotation 
\begin{equation}
 J =  \pm \int_0^\infty \frac{dp}{2\pi}  \left[ \frac{2 p}{1+ 
e^{\frac{p-\mu  }{T}} } + \frac{2 p}{1+ e^{\frac{p+\mu  }{T}} }\right]  = \pm 
\left( \frac{\mu  _{L,R}^2}{2\pi} + \frac{\pi T^2}{6}\right) \,.
\end{equation}
and the energy current
\begin{equation}
 J_{\epsilon} =  \pm \int_0^\infty \frac{dp}{2\pi}  \left[ \frac{2 p^2}{1+ 
e^{\frac{p-\mu  }{T}} } - \frac{2 p^2}{1+ e^{\frac{p+\mu  }{T}} }\right]  = 
\pm \left( \frac{\mu  _{L,R}^3}{3\pi} + \frac{\mu  _{R,L} \pi T^2}{3}\right) \,.
\end{equation}
These (admittedly somewhat hand-waving) arguments give rise to the chiral 
vortical effects (CVE) in the current and energy current
\begin{align}\label{eq:cvechiral}
 \vec J &= \pm \left( \frac{\mu  _{L,R}^2}{4\pi^2} + \frac{T^2}{12} \right) \vec 
\omega \,\\ \label{eq:cveenergy}
  \vec J_\epsilon & = \pm \left( \frac{\mu  _{L,R}^3}{6\pi^2} + \frac{\mu  
_{L,R} T^2}{6} \right) \vec \omega \,
\end{align}
Very nice! We get currents from the lowest Landau level. This is quite interesting 
by itself but it gets even a bit more interesting
remembering that the lowest Landau level was also responsible for the anomaly. 
So is there a more direct connection? And this answer is yes! As a first step 
let us 
generalize our results for single fermions to many different species of fermions 
and label them by an index $f$ (we can also switch to a basis with left-handed 
fermion only).
Then assume that there is a bunch of $U(1)$ symmetries labeled by an index $a$ 
under which the fermions carry charges $q^f_a$. The chemical potential for the 
fermion
species $f$ is then $\sum_a q^f_a \mu  _a = \mu  ^f$ and it sees the magnetic 
field $\vec B^f = \sum_a q_a^f \vec B_a$. The current corresponding to symmetry 
$a$ is likewise
$J_a = \sum_f q^f_a J^f$. Now we use the expression for the elementary chiral 
currents and obtain (summation over repeated indices is implied here)
\begin{empheq}[box=\fbox]{align}\label{eq:cmegeneral}
 \vec J_a &=  d_{abc} \frac{\mu  _b}{4\pi^2} \vec B_c\,, & ~ & 
 \vec J_\epsilon = \left( d_{abc} \frac{\mu  _a \mu  _b}{8\pi^2} + b_c 
\frac{T^2}{24} \right)\vec B_c \, \,\\
\label{eq:cvegeneral}
\vec J_a &= \left( d_{abc} \frac{\mu  _b \mu  _c}{4\pi^2} + b_a \frac{T^2}{12} 
\right)\vec \omega \,,&~&
  \vec J_\epsilon = \left( d_{abc} \frac{\mu  _a \mu  _b \mu  _c}{6\pi^2} + b_a 
\mu  _a \frac{T^2}{6} \right)\vec \omega  \,.
\end{empheq} 
Where $d_{abc}$ and $b_a = \sum_f q^f_a$ are the
chiral and gravitational anomaly coefficients \cite{Landsteiner:2011cp}. 
This is a hint towards the deep origin of these expression. The currents are 
indeed induced by the chiral and gravitational anomaly as we will discuss
in the next sections. We have 
obtained these expressions without having to worry about regularization issues. 
The
integrals are finite because they are damped in the UV by the exponential decay 
of the distribution functions. Furthermore in view of what we learned about the 
spectral
flow we also know that naive counting of particles around the Fermi surface (or 
the normal ordered vacuum) gives us only the covariant current. To get the consistent
current additional terms can and do arise from the Bardeen-Zumino polynomials.

\begin{figure}[ht]
\includegraphics[width=\textwidth]{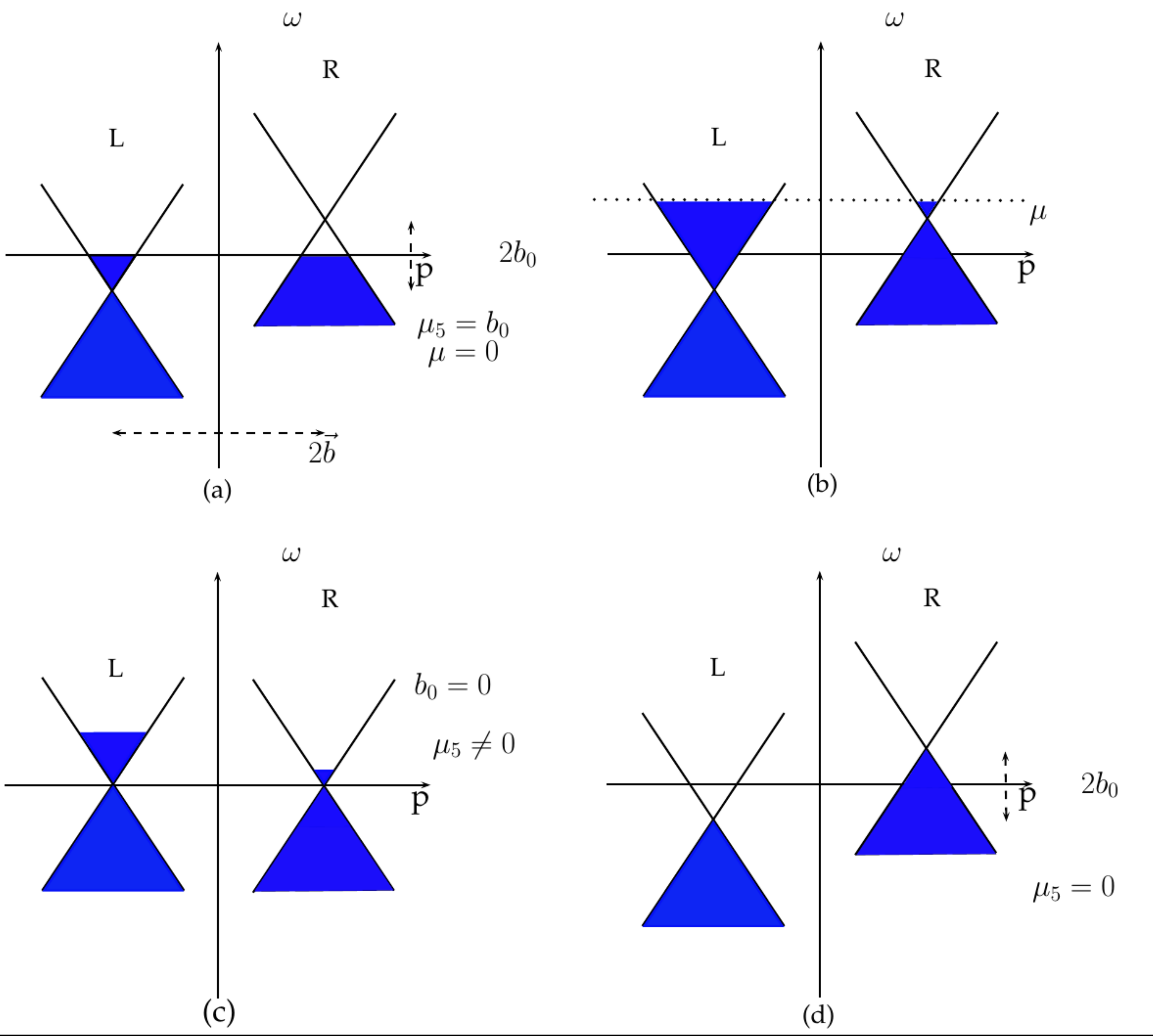}
\caption{The figure shows different fillings corresponding to different chemical 
potentials and axial vector field backgrounds 
(with notation $A_0^5=b_0$, $\vec A_5 = \vec b$). In panel (a) $\mu  _5=A_0^5$ 
and $\mu  =0$ so neither CME nor CSE are present.
In panel (b) still $\mu  _5=A_0^5$ but now $\mu  >0$ so only the CME vanishes. 
In panel (c) we have $\mu  _5\neq 0$ and $\mu  \neq 0$ but
$A_0^5=0$ so now both CME and CSE are non-vanishing. Finally in panel (d) both 
chemical potentials are zero but $A_0^5\neq 0$ so
here only the CME is non-vanishing. }
\label{fig:weylcones}
\end{figure}

Let us now specialize to the case of one Dirac fermion: we have 
$d_{5VV}=d_{V5V}=d_{VV5}=2$ , $d_{555}=2$ and $b_5=2$. Let us study the magnetic 
effects and include
also the Chern-Simons current from the Bardeen-Zumino polynomial, i.e. we want 
to write down the magnetic field induced currents in terms of the conserved 
consistent current.
\begin{empheq}[box=\fbox]{align}\label{eq:cmeproper}
 \vec{\mathcal{J}} &= \frac{\mu  _5-A_0^5}{2\pi^2} \vec{B}\,,\\ \label{eq:cse}
 \vec{\mathcal{J}}_5 & = \frac{\mu  }{2\pi^2} \vec{B}\,.
\end{empheq}
These two are are the (proper) chiral magnetic effect and the chiral separation 
effect (CSE). While they look very similar if written down in terms of the 
covariant current
they are quite different once written in terms of the consistent current. There 
is no contribution to the CSE by any Bardeen-Zumino polynomial. Ultimately this 
can be traced
back to the fact that it depends only on the chemical potential $\mu  $ related 
to a truly conserved current. On the other hand the CME receives two 
contributions, one from
the states that fill the energy range between the vacuum and the Fermi surface 
and another one that stems from the boundary conditions in momentum space and 
depends
on the specific gauge invariant regularization.  We have already mentioned that 
the role of $A_0^5$ is to shift the tips of the Weyl cones in opposite directions 
in energy. If this
shift is equal to the imbalance in the chemical potentials then there is a 
perfect cancellation between the two terms and the CME vanishes. In fact if one 
defines the
grand canonical ensemble for axial charge in the usual way by introducing a 
Hamiltonian $H - \mu  _5 Q_5$ this has the effect of achieving both effects:
a non-vanishing occupation number but also a shift in the locations of the tips 
of the Weyl cones in precisely such a way that $A_0^5=\mu  _5$. This definition 
of 
equilibrium grand canonical ensemble results in a vanishing CME! 
The different situations of vanishing vs. non-vanishing CME are sketched in 
figure \ref{fig:weylcones}.

\section{Hydrodynamics and triangle anomalies}

So far we have derived the CME and the CVE (with some hand waving) only for free 
fermions. We need to convince ourselves that the CVE expressions have some
physical meaning. We could try to do better and study an ensemble of rotating
fermions, except that this is still rather ill-defined. In a relativistic theory 
an equilibrium 
ensemble with constant rotation cannot exist since at some large enough radius 
the tangential velocity necessarily would exceed the speed of light. 
In addition we need to include interactions as well. The lowest Landau level is 
subject to an index theorem that protects it, so we might expect that 
interactions don't do
modify the expressions for the CME.  An effective way of studying an interacting 
gas or more general a fluid of chiral fermions is the formalism of 
hydrodynamics.
Hydrodynamics is an effective field theory describing the time evolution of 
systems near equilibrium\footnote{Although demanding closeness to 
equilibrium is probably too restricted. A modern point of view is that
systems even relatively far from equilibrium can be well described by
hydrodynamics \cite{Heller:2016gbp} }. 
One assumes that locally the system equilibrates fast
such that local versions of thermodynamic parameters, the temperature $T$ and 
the chemical potential $\mu  $ can be defined. Intuitively thermodynamics is 
what is left of a system if we forget all the details that can be forgotten. 
That means we know nothing but the indestructible conserved charges, like energy 
and
$U(1)$ charges. Now we promote these to local notions of energy density and 
charge density. In relativistic systems we therefore need a four vector $J^\mu  
$ to
describe the dynamics of the $U(1)$ charge and a symmetric second rank tensor 
$T^{\mu  \nu}$ to describe the dynamics of the energy and momentum densities. 
Let us study energy-momentum conservation in quantum field theory. We assume a 
quantum action depending on a source field coupled to the
energy momentum tensor. This source field is the metric. If there are no 
diffeomorphism anomalies this quantum action should be diffeomorphism invariant
such that
\begin{equation}
 \int d^4x\, \sqrt{-g} \left[\frac{2}{\sqrt{-g}} \frac{\delta \Gamma}{\delta 
g_{\mu  \nu}} \nabla_\mu  \epsilon_\nu + \frac{1}{\sqrt{-g}} \frac{\delta 
\Gamma}{\delta A_\mu  }\left( \partial_\mu  (\xi^\lambda A_\lambda) + 
\xi^\lambda F_{\lambda\mu  } \right) \right] =0\,.
\end{equation}
Where we used that the metric and gauge field transform under diffeomorphisms 
as 
\begin{align}
 \delta g_{\mu  \nu} &= \nabla_\mu  \xi_\nu + \nabla_\nu \xi_\nu\,,\\
 \delta A_\mu  &= \partial_\mu  (\xi^\lambda A_\lambda) + \xi^\lambda 
F_{\lambda\mu  }\,,
\end{align}
where the vector field $\xi^\mu  $ is the generator of the diffeomorphism. 
Furthermore we identify 
$\frac{2}{\sqrt{-g}} \frac{ \delta \Gamma }{ \delta g_{\mu  \nu} } = T^{\mu  
\nu}$ and $\frac{1}{\sqrt{-g}} \frac{\delta \Gamma}{\delta A_\mu  } = 
\mathcal{J}^\mu  $.
Since $\xi^\lambda$ is arbitrary we infer that the energy momentum tensor has to 
fulfill 
\begin{equation} \label{eq:emconservationcons}
 \nabla_\mu   T^{\mu  \nu} = F^\nu\,\mu  \mathcal{J}^\mu  - A^\nu 
\nabla_\mu  \mathcal{J}^\mu  \,.
\end{equation}
So the energy-momentum 
tensor is naturally symmetric and conserved up to the terms on the
right hand side. The first term represents the injection of energy and momentum 
by the external electric and magnetic fields. The second term vanishes
if the current is conserved. When an anomaly is present it does 
contribute to the energy momentum conservation. 
Now we use our knowledge about anomalies gained in the 
previous sections. They come as consistent and as covariant
ones and the covariant one couples only to the external field strengths but not 
to the external vector field $A_\mu$. Indeed if we write the consistent
current as covariant current plus Bardeen-Zumino polynomial the term 
proportional to the anomaly in (\ref{eq:emconservationcons}) vanishes.
So energy-momentum and charge conservation laws in terms of the covariant 
current takes the simpler form
\begin{align} \label{eq:hydroeoms}
 \partial_\mu   T^{\mu  \nu} & = F^\nu\,_\mu   J^\mu  \,, \\
 \partial_\mu   J^\mu  &= \frac{1}{32\pi^2} \epsilon^{\mu  \nu\rho\lambda} 
F_{\mu  \nu} F_{\rho\lambda}\,,
\end{align}
where we have also set the metric to the flat one.
These are already the equations of motion for hydrodynamic effective field 
theory. The five equations can determine five dynamical variables which are 
commonly
parametrized as local temperature $T$, chemical potential $\mu  $ and the fluid 
velocity $u^\mu  $ obeying $u^2=1$.  Energy-momentum tensor and charge current 
are now expressed 
in terms of the local fluid variables in a derivative expansion (see 
\cite{Kovtun:2012rj} for a recent review of relativistic hydrodynamics)
\begin{align}
 T^{\mu  \nu} &= T^{\mu  \nu}_{(0)} + T^{\mu  \nu}_{(1)} + \dots\,\\
 J^\mu  &= J^\mu  _{(0)} + J^\mu  _{(1)} + \dots\,.
\end{align}
To zeroth order in derivatives 
\begin{align}
 T^{\mu  \nu}_{(0)} & = (\varepsilon + p) u^\mu   u^\nu - p\, \eta^{\mu  \nu} 
\,.\\
  J^\mu  _{(0)} &= \rho\, u^\mu  \,.
\end{align}
The energy density, pressure and charge density are computed via thermodynamic 
relations from the free energy $F(T,\mu  )$ defined at equilibrium. 

To first order in derivatives one has to deal with a large set of ambiguities 
since redefinitions of the local temperature, chemical potential and fluid 
velocity 
of the form $T\rightarrow T+\delta T$ etc. compete with the terms at higher 
order in derivatives. These ambiguities are fixed by a choice of frame. One
of the standard frames is the so-called Landau frame, which is defined by 
demanding $T^{\mu  \nu} u_\mu  = \varepsilon u^\mu  $ to all orders. 
It is also useful to define the projector transverse to the fluid velocity 
$P^{\mu  \nu} = \eta^{\mu  \nu} - u^\mu   u^\nu$. 
Another important point is that for the covariant current only covariant field 
strenghts can enter the expression. It is useful to introduce electric and
magnetic fields in the local rest frame $B^\nu = \frac{1}{2} \epsilon^{\mu  
\nu\rho\lambda} u_\mu   F_{\rho\lambda}$ and $E^\mu  = F^{\mu  \nu} u_\nu$ and
the local vorticity $\omega^\nu = \frac{1}{2} \epsilon^{\mu  \nu\rho\lambda} 
u_\mu  \partial_\rho u_\lambda$. 
The first order terms then take the form
\begin{align}\label{eq:hydroonederivativeT}
 T^{\mu  \nu}_{(1)} &=  -\eta P^{\mu  \alpha} P^{\nu\beta} \left( 
\partial_\alpha u_\beta + \partial_\beta u_\alpha - 2/3 \eta_{\alpha\beta} 
\partial_\lambda u^\lambda \right) - 
		    \zeta P^{\alpha\beta}  \partial_\lambda  u^\lambda\,,\\ 
\label{eq:hydroonederivativeJ}
 J^{\mu  }_{(1)} & = -  \sigma T P^{\mu  \alpha} \partial_\alpha \left( 
\frac{\mu  }{T} \right) + \sigma E^\mu  + \xi_B B^\mu  + \xi_\omega \omega^\mu  
\,.
\end{align}
Here $\eta$ is the shear viscosity, $\zeta$ is the bulk viscosity and $\sigma$ is 
the electric conductivity. The new coefficients $\xi_B$ and $\xi_\omega$ give
room for response to external magnetic and electric fields. 
One more important point about hydrodynamics is that it also includes a local 
form of the second law of thermodynamics. One needs to define an entropy
current with a non-negative divergence. Again the entropy current can be built 
up in a derivative expansion. It takes the form
\begin{equation}
 S^\mu  = s u^\mu  - \frac{\mu  }{T} J^\mu  _{(1)} + D_B B^\mu  + D_\omega 
\omega^\mu  \,,
\end{equation}
where $s$ is the local entropy density and $D_B$ and $D_\omega$ new response
coefficients. 
Now one demands
\begin{equation}\label{eq:secondlaw}
 \partial_\mu   S^\mu  \geq 0
\end{equation}
and this constrains the the transport coefficients in such way that $\eta, 
\zeta, \sigma \geq 0$. Surprisingly this constraint is much stronger for 
response coefficients due to magnetic field or rotation. In  \cite{Son:2009tf, 
Neiman:2010zi} it was shown that these coefficients are almost completely 
determined
by the second law (\ref{eq:secondlaw}) up to some undetermined integration 
constant $\gamma$
\begin{align}
 \xi_B &= \frac{1}{4\pi^2} \left( \mu  - \frac{1}{2} \frac{\rho}{\varepsilon+p}  
(\mu  ^2 +  \gamma T^2) \right) \,\\
 \xi_\omega & =  \frac{1}{4\pi^2} \left( \mu  ^2 + \gamma T^2 - \frac{2}{3} 
\frac{\rho}{\varepsilon+p} \left( \mu  ^3 + 3 \gamma \mu   T^2 \right) \right) 
\,,\\
D_B &= \frac{1}{8\pi^2} \frac{ \mu  ^2}{T} + \gamma T \,,\\
D_\omega & = \frac{1}{12\pi^2} \frac{\mu  ^3}{T} + 2\gamma \mu   T \,.
\end{align}
Furthermore the new coefficients do not contribute to entropy production, i.e.
they describe dissipationless transport. 
If we compare these with the expressions (\ref{eq:cmegeneral}), 
(\ref{eq:cvegeneral}) we see some similarity but also differences, the most 
striking is that
there is no energy current. This can be traced back to the underlying choice of 
frame. We can go to another frame by redefining the four velocity such that the
terms depending on charge $\rho$, energy $\epsilon$ and pressure $p$ vanish. 
This can be done by redefining $u^\mu  \rightarrow u^\mu  + \delta u^\mu  $ where
\begin{equation}
 \delta u^\mu  = \frac{1}{8\pi^2} \frac{1}{\varepsilon+p}\left[ (\mu  ^2 + \gamma T^2) 
B^\mu  + \frac{4}{3} (\mu  ^3 + 3 \gamma \mu   T^2) \omega^\mu   \right] \,. 
\end{equation}
In this frame the anomaly related transport coefficients take the form 
\cite{Amado:2011zx}
\begin{align}
 T^{\mu  \nu} & = \sigma_{\varepsilon,B} ( u^\mu   B^\nu + u^\nu B^\mu  ) +  
\sigma_{\varepsilon,\omega} ( u^\mu  \omega^\nu + u^\nu \omega^\mu  ) \,.\\
 J^\mu  &= \sigma_{B} B^\mu  + \sigma_\omega \omega\,,
\end{align}
If we expand for small velocities $u^\mu  = (1,\vec v)$ to first order in $\vec 
v$ we can identify the $\sigma$ coefficients with (\ref{eq:cmechiral}),
(\ref{eq:cmeenergy}), 
(\ref{eq:cvechiral}), (\ref{eq:cveenergy}).
This also fixes the so far undetermined constant $\gamma=\frac{\pi^2}{3}$.

The frame has a very nice physical interpretation \cite{Stephanov:2015roa}. In 
the presence of anomalies the chiral magnetic and chiral vortical effects give 
rise
to dissipationless  charge and energy flows.  In the new frame the four velocity 
$u^\mu  $ parametrizes then only the dissipative ``normal'' flow.
If one allows momentum to relax e.g by placing an obstacle such as a heavy 
impurity into the flow then on long time scales all the normal
flow will vanish described by $u^\mu  =0$ while charge and momentum flow induced 
by CME and CVE are still present.

An alternative way of deriving the transport coefficients of a fluid subject
to anomalies is the effective action approach \cite{Haehl:2013hoa}. Higher orders in derivatives
have been studied in \cite{Kharzeev:2011ds,Megias:2013joa}.

The entropy argument fixes the dependence of the anomalous transport 
coefficients on the chemical potentials but not their temperature
dependence. On the other hand we have seen already that the temperature 
dependence naturally has the gravitational anomaly coefficient
attached to it. Since the gravitational anomaly is fourth order in derivatives 
and hydrodynamics is developed only to the one derivative level here
there is no natural way it can enter here. 
A more general argument based on geometry was presented in \cite{Jensen:2012kj}
and a consistency condition based on the global gravitational anomaly was put forward
in \cite{Golkar:2015oxw}.
Probably the most straightforward way of demonstrating a direct
relationship between the gravitational anomaly and the temperature dependence of 
anomalous transport coefficients comes from holography as we will review below.

\section{Holography}
The AdS/CFT correspondence or holographic correspondence 
\cite{Maldacena:1997re} has developed over the last years into a very useful and 
powerful
tool for studying strongly coupled field theories at finite temperature and 
density. Recent useful reviews are \cite{Zaanen:2015oix, Ammon:2015wua}

Before discussing  how holography can be used to gain insight into anomalous 
transport  we very briefly review the basics of the AdS/CFT correspondence.

The origin of the AdS/CFT correspondence is the duality between Type IIB string 
theory on $\mathrm{AdS}_5 \times \mathrm{S}_5$ and ${\cal N}=4$ supersymmetric gauge theory. 
The ${\cal N}=4$ supersymmetric gauge theory is 
a non-abelian, four dimensional quantum field theory whose field content 
consists of six scalars, four Majorana fermions and a
vector field. They all transform under the adjoint representation of the gauge 
group $SU(N)$.  It features four supersymmetries and this fixes all the 
couplings between the different fields. As it is a gauge theory physical 
observables are gauge invariant
operators such as $\mathrm{tr}(F_{\mu  \nu} F^{\mu  \nu})$. The global symmetry 
group $SO(6)$ acts on the scalars and the
fermions (in the $SU(4)$ spin representation of $SO(6)$).

The dual theory is a theory of gravity (this is what type IIB string theory is) 
but living in quite a few more dimensions, ten as
opposed to the four the field theory knows about. But five of these ten are 
easily got rid off: the isometries of the $S_5$ part
of the geometry form $SO(6)$. The $S_5$ is the geometric realization of what 
appears as an internal, global symmetry group in the field theory. 
 
The field theory is characterized by two parameters, the gauge coupling $g_{YM}$ 
and the rank of the gauge group $N$. The dual 
string theory has a string coupling $g_s$ (the amplitude for a string to split 
in two) and a fundamental length scale $l_s$, the string scale. 
Furthermore the geometry is determined by a scale $L$ determining the curvature 
of the AdS$_5$ as $R=-20/L^2$. The AdS/CFT correspondence relates these 
parameters
in the following way:
\begin{eqnarray} \label{eq:holodictionary}
g^2_{YM} N &\propto& \frac{L^4}{l_s^4}\\
1/N &\propto& g_s
\end{eqnarray}
The AdS/CFT correspondence is therefore a strong weak coupling duality: for weak 
curvature we have large $L$ and therefore also
large 't-Hooft coupling. In this regime of weak curvature stringy effects are 
negligible and we can approximate the string theory
by type IIB supergravity. If we furthermore take the rank of the gauge group $N$ 
to be very large we can also neglect quantum
loop effects and end up with classical supergravity! This is the form of the 
correspondence most useful for the applications 
to many body physics: classical (super)gravity on $(d+1)$ dimensional Anti-de 
Sitter space.

Based on this  original example we can conjecture that every theory of gravity 
plus some suitably chosen matter fields on $(d+1)$ dimensional Anti-de Sitter 
space is a dual to a certain quantum field theory in $d$ dimensions. In fact we 
might even be a bit more brave and delete the words ``dual to a'' in the 
previous phrase. This is the point of view taken in the applications of the 
AdS/CFT correspondence to the world of solid state physics. The additional 
matter fields are then chosen to reflect a particular symmetry content of the 
underlying quantum field theoretical system one is interested in. 
Having this in mind we will forget from now on some of the seemingly non-essential\footnote{
Note however that non-supersymmetric AdS solutions to string theory have been conjectured to
be unstable recently \cite{Ooguri:2016pdq}.}
ingredients, such as supersymmetry, 
string theory and extra dimensions in form of the 
$S_5$.

For the applications to quantum field theory it is most useful to write the $AdS$ 
metric in form
\begin{equation}\label{eq:ads}
 ds^2 = \frac{r^2}{L^2} ( -dt^2 + d\vec{x}^2 ) + \frac{L^2}{r^2}dr^2\,.
\end{equation}
The space on which the dual quantum field theory lives is recovered by taking 
the limit 
$ds^2_\mathrm{QFT} = \lim_{r\rightarrow \infty} r^{-2} ds^2$. This is why 
sometimes it is said that the dual field theory
lives on the ``boundary'' of AdS and why the AdS/CFT correspondence is also 
referred to as ``Holography``. The physical interpretation of the holographic
direction is that it represents an energy scale. We can identify the 
high-energy UV limit in the field theory with the
$r\rightarrow \infty$ limit in the AdS geometry, whereas the low-energy IR limit 
is $r\rightarrow 0$. 

The asymptotic behavior of the fields in AdS has the form 
\begin{equation}\label{eq:asymptoticexpansio}
 \Phi = r^{-\Delta_-} \left(\Phi_0(x) + O(r^{-2})\right) + r^{-\Delta_+} 
\left(\Phi_1(x) + O(r^{-2} )\right)\,.
\end{equation}
The exponents $\Delta_{\pm }$ obey $\Delta_- < \Delta_+$ and depend on the 
nature of the field, e.g. for a scalar field of 
mass $m$ they are $\Delta_{\pm } = \frac 1 2 (d\pm \sqrt{d^2+4m^2L^2})$. 

We now would have to evaluate the path integral over
the fields in AdS keeping the boundary values $\Phi_0(x)$ fixed. The result is a 
functional depending on the boundary
fields $\Phi_0(x)$. Now the boundary field $\Phi_0(x)$ is interpreted as the 
source $J(x)$ that couples to a (gauge invariant)
operator ${\cal O}(x)$ of conformal dimension $\Delta_+$ in the field theory
\begin{equation}
 Z[J] = \int_{\Phi_0 = J} d \Phi\; \exp(-i S[\Phi])
\end{equation}
Connected Green's functions of gauge invariant operators in the quantum field 
theory can now be generated by functional
differentiation with respect to the sources
\begin{equation}
\left\langle {\cal O}_1(x_1) \cdots {\cal O}_n(x_n) \right\rangle = \frac{\delta^n 
\log Z}{\delta J_1(x_1) \cdots\delta J_n(x_n)}\,.
\end{equation}
In the limit in which the gravity theory becomes classical, i.e. the large $N$ 
and large coupling $g^2_{YM} N$ limit the
path integral is dominated by the classical solutions to the field equations and 
$\log Z$ can be replaced by the
classical action evaluated on a solution of the field equations. In this case 
the coefficient $\Phi_1(x)$ of the asymptotic
expansion (\ref{eq:asymptoticexpansio}) is the vacuum expectation value of the 
operator sourced by $\Phi_0$.
\begin{equation}
 \left\langle {\cal O}(x)\right\rangle \propto \Phi_1(x)\,.
\end{equation}

To explicitly compute $\Phi_1(x)$ we need to supply a second boundary condition, 
so far we have fixed only the asymptotic value
$\Phi_0$. For time independent solutions we demand regularity in the interior of 
the (possibly only asymptotically) AdS space.
For time dependent perturbation one needs to impose the more general 
''infalling`` boundary conditions.

The holographic dictionary relates in this way (gauge-invariant) local 
operators to fields in the bulk of AdS and can be summarized
in table \ref{tab:dictionary}

\begin{table}
\begin{center}
\begin{tabular}{|c|c|}\hline
 Gravity &  QFT \\  \hline \hline
 $D+1$ dimensions & $D$ dimensions \\ \hline
 holographic direction & RG scale \\ \hline
 strongly coupled & weakly coupled \\ \hline
 scalar field & scalar operator \\ \hline
 gauge field & conserved current \\ \hline
 metric tensor &  energy-momentum tensor \\ \hline
\end{tabular}
\caption{The holographic dictionary}\label{tab:dictionary}
\end{center}
\end{table}

Anomalies can be incorporated rather easily: they are represented by a five 
dimensional Chern-Simons term. So to study anomalous transport with the
means of holography the following model is a good starting point 
\cite{Landsteiner:2011iq}
\begin{align}\label{eq:holomodelanomal}
 S &=  \int d^5x\,\sqrt{-g} \left[
R + 2 \Lambda - \frac{1}{4} F_{MN} F^{MN}  +   \right.    \nonumber \\
 & \left. \epsilon^{MNPQR} A_M  \left( \frac{\alpha}{3} F_{NP} F_{QR} + \lambda  
R^A\,_{BNP} R^B\,_{AQR} 
\right) 
\right]\,.
\end{align}
The model contains only one five dimensional $U(1)$ gauge field. In the dual 
field theory we consider only one anomalous symmetry. The anomaly is 
of $U(1)^3$ type, of a single chiral fermion and the mixed chiral-gravitational 
anomaly. We note that the action (\ref{eq:holomodelanomal}) has diffeomorphism 
symmetry
and that the $U(1)$ symmetry is preserved up to a boundary term. This boundary 
term takes precisely the form of the chiral and chiral-gravitational anomaly. 
We can also derive the operators $J^\mu  $ and $T^{\mu  \nu}$ as the variations 
of the on-shell action with respect to the boundary values of the gauge field 
\cite{Megias:2013joa}
and the metric. 
\begin{align}\label{eq:consholoJ}
 \mathcal{J}^\mu  &= \sqrt{-g} \left(  F^{r\mu  } + \frac{4\alpha}{3} 
\epsilon^{\mu  \nu\rho\lambda }A_\nu F_{\rho\lambda}  \right) \,\\ 
\label{eq:consholoT}
T^{\mu  \nu}& = \sqrt{-g} \left( \frac 1 2 K^{\mu  \nu} - \frac{1}{2} K 
\gamma^{\mu  \nu} +
2 \lambda \epsilon^{(\mu  \rho\lambda\sigma}\left( 
\frac{1}{2} F_{\rho\lambda} \hat R^\nu)_\sigma + D_\delta ( A_\rho \hat 
R^{\delta\nu)}\,_{\lambda\sigma}) 
\right)
\right)
\end{align}
Here greek indices are four dimensional boundary indices, $\hat R$ the intrinsic 
four dimensional curvature of the boundary, $\gamma_{\mu  \nu}$ the
boundary metric, $D_\mu  $ the covariant (with respect other $\gamma_{\mu  \nu}$) 
boundary derivative. The divergences are
 \begin{align}\label{eq:holoanomaly}
D_\mu  \mathcal{J}^\mu   &= \epsilon^{\mu  \nu\rho\lambda} \left( 
\frac{\kappa}{3} F_{\mu  \nu} F_{\rho\lambda} + \lambda \hat 
R^{\alpha}\,_{\beta\mu  \nu} \hat R^{\beta}\,_{\alpha\rho\lambda}\right) \,,\\
D_\mu   T^{\mu  \nu}  &=   F^{\nu\mu  } \mathcal{J}_\mu  + A^\nu D_\mu  \mathcal{J}^\mu  \,,
\end{align}
which are nothing but the consistent form of the chiral and chiral-gravitational 
anomalies.

Now we have our holographic theory but need a state with chemical potential $\mu 
 $ and temperature $T$. This is represented by a charged black hole solution.
The five dimensional metric takes the form
\begin{equation}
 ds^2 = \frac{dr^2}{r^2 f(r)} - r^2 f(r) dt^2 + r^2 d\vec{x}^2\,.
\end{equation}
We assume that $\lim_{r\rightarrow \infty}f(r) = 1$ (the spacetime is 
asymptotically AdS)  and that at some finite value $f(r_H)=0$ there is a 
non-degenerate horizon with temperature
\begin{equation}
 T = \frac{ r_H^2 f'(r_H) }{4 \pi}\,.
\end{equation}
The chemical potential is related to a non-trivial profile of the temporal 
component of the gauge field. In its most elementary definition the chemical 
potential is the
energy that is need to add one unit of charge to the thermal ensemble. The 
thermal ensemble is represented by the black hole and to add a unit of charge is 
to take 
this charge from infinity to behind the horizon. The chemical potential can 
therefore be identified with the difference of the potential energy at the 
boundary and at the
horizon
\begin{equation}
 \mu  =  A_0(\infty) - A_0(r_H)\,.
\end{equation}
A subtlety related to the value of the gauge field on the 
horizon \cite{Gynther:2010ed} should be mentioned. If one defines the equilibrium 
state as the one with a smooth Euclidean geometry then
we need also to impose $A_0(r_H)=0$. This constraint comes from the fact that 
the Euclidean time $t=i\tau$ the black hole geometry simply ends at $r=r_h$, 
there is 
no interior and the geometry is smooth only if $\tau$ is periodic with 
periodicity $1/T$.
 If $A_0(r_H)\neq 0$ an integral $\oint A_0 d\tau$ would be non vanishing for an 
infinitely small circle and indicate a field strength of delta function 
support at $r=r_H$.

The equations of motion for $f(r)$ and $A_0(r)$ are
\begin{align}
 (r^3 A'_0(r))' &=0 \,,\\
 f'(r) +\frac{4}{r} f(r) - \frac{4}{r} + \frac{ (A'_0(r))^2}{6r}&=0\,.
\end{align}
The solution for the gauge field is $A_0(r) = \alpha - \frac{\rho}{2r^2}$. The 
chemical potential is  $\rho = 2 r_H^2 \mu  $.
If we impose vanishing of the gauge field at the horizon we also have 
$\alpha=\mu  $.
The solution for the metric component is $f(r) = 1 - \frac{M}{r^4} + 
\frac{\rho^2}{12 r^6}$. 

Before we go on and compute the chiral magnetic and chiral vortical effects we 
note that the current (\ref{eq:consholoJ})  energy-momentum tensor (\ref{eq:consholoT}) have a trivial
part that is determined solely by the boundary values of the fields. Since we 
want to have a flat boundary metric these terms vanish for the energy momentum
tensor. For the current the corresponding Chern-Simons term is just the 
Bardeen-Zumino polynomial that relates covariant and consistent current. 
In holography the UV origin of the Bardeen-Zumino polynomial is manifes as 
it is a pure boundary term.
We conclude that the holographic form of the covariant current is 
\begin{equation}
 J^\mu  _\mathrm{cov} = \sqrt{-g} F^{r\mu  } |_{r\rightarrow \infty}
\end{equation}

Now we are ready to compute chiral magnetic and chiral vortical effects. In 
order to do so we introduce appropriate sources. For the magnetic field
this is straightforward we need to impose the boundary condition
\begin{equation}
 A_y|_{r=\infty} = B x
\end{equation}
or any equivalent gauge. 

For the chiral vortical effect we need to implement boundary conditions  
encoding rotation. That can be done via a metric perturbation. We remember
that metric components with mixed time and space indices represent metric fields 
generated by rotating bodies and they induce by themselves rotation
via the frame dragging effect. In fact in the formalism of gravito 
electromagnetism rotation is represented via the gravito-magnetic field 
$\mathcal{B}_i = \epsilon_{ijk} \partial_j g_{0k}$. A simple way of seeing this is via a 
fluid picture. The fluid at rest can be defined by the 
contra-variant four vector $u^\mu  = (1,0,0,0)$ while the vorticity is defined 
through the co-variant components $\omega^\mu  = \frac 1 2 \epsilon^{\mu  
\nu\rho\lambda} u_\nu \partial_\rho u_\lambda$.
Therefore for small velocities vorticity and gravito-magnetic field are related 
as  $ 2 \vec\omega = \vec{\mathcal{B}}  $

 This motivates us to chose the metric perturbations
\begin{equation}
 g_{ty} = \mathcal{B} x f(r) r^2\,.
\end{equation}
It turns out that the gravito magnetic field also induces a gauge field 
\begin{equation}
 A_y =\mathcal{B} x \frac{\rho}{2r^2}
\end{equation}

Now the linear response due to $B$ and $\mathcal{B}$ can be computed via the equations 
of motion. We will concentrate on the response in the current. The corresponding
equation of motion is 
\begin{equation}\label{eq:eomaz}
 (r^3 f(r) a'_z(r) )'    - \frac{\rho}{2} h'^t_z(r)     -  8  \alpha B\frac{ 
\rho}{r^3}  - \alpha\mathcal{B} \frac{4\rho^2}{r^5}  + \lambda \mathcal{B} [ 2 
r^4 f'(r)^2]' =0 \,,
\end{equation}
where a prime denotes $\partial/\partial r$.
We note that the covariant current in the $z$ direction is $J_z = 
\lim_{r\rightarrow\infty}r^3 f a'_z$. All the terms in this equation are total 
derivatives.
This tell us that 
\begin{equation}
  (r^3 f(r) a'_z(r) )    - \frac{\rho}{2} h^t_z(r)     + 4 \alpha B 
\frac{\rho}{r^2}  + \alpha\mathcal{B} \frac{\rho^2}{r^4}  + \lambda\mathcal{B}  
2 r^4 f'(r)^2 = \mathrm{const}.
\end{equation}
Evaluating this on the horizon assuming regularity of $a_z$ there and demanding 
also that the metric perturbation $h^t_z$ vanishes at the horizon allows us to 
compute the 
constant on the right hand side. Then we can evaluate this equation on the 
boundary where all terms except the constant on the right hand side and
the current vanish (the metric perturbation has to fall off at least as 
$1/r^4$). This gives the result
\begin{equation}
 J_z = 8\alpha \mu   B_z + (4\alpha \mu  ^2 + 32\pi^2 \lambda 
T^2)\mathcal{B}_z\,.
\end{equation}
Comparing (\ref{eq:holoanomaly}) to the anomalies of a single chiral fermion we 
can identify $\alpha=\frac{1}{32\pi^2}$ 
and $\lambda=\frac{1}{768\pi^2}$.  Then these are the same results as in the free field theory! 
The relation 
between the chiral-gravitational anomaly and the temperature dependence of the 
chiral vortical
effect is very direct in holography. It is interesting to note how holography 
manages to correct the mismatch in derivatives. It is still true of course that 
the gravitational
anomaly and also the five dimensional gravitational Chern-Simons term are higher 
order in derivatives. And as can be seen in equation (\ref{eq:eomaz}) the 
term proportional to $\lambda$ is indeed of fourth order. But the trick of 
holography is that three of these derivatives are along the holographic 
direction and only
one is along the actual spacetime!
Non-renormalization of the CME and CVE have been discussed in holography in
\cite{Gursoy:2014boa, Gursoy:2014ela,Grozdanov:2016ala}, using field theory 
in \cite{Golkar:2012kb, Hou:2012xg}, in a model 
independent way based on a mixed hydrodynamic/geometric approach \cite{Jensen:2012kj} and a 
more general relation to gravitational anomalies has also been investigated in 
\cite{Golkar:2015oxw,Chowdhury:2016cmh}.

We need to discuss one more point: we have set the horizon value of the metric 
perturbation $h^t_z(r)$ to zero. More generally
we can leave it arbitrary and fix it later. It turns out that this integration 
constant corresponds to the choice of frame that we discussed in hydrodynamics, 
e.g. we could
have used it to set the response in the energy-momentum tensor to zero, then we 
could recover the expression of chiral vortical and chiral magnetic effect 
in Landau frame.


\section{Anomalous transport and  Weyl semi-metals}

As stated in the introduction anomalous transport phenomena play an important role 
in many branches of 
physics.
One of the most interesting and active fields of applications of anomalous 
transport phenomena is condensed matter theory and experiment.
The electronics of a new class of materials known as Dirac- and Weyl 
(semi-) metals \cite{WSM1,WSMreviewTV,burkovbalents1} (or more generally topological
metals) is governed by the Weyl equation. 
In this chapter we will briefly review the physics of Weyl semi-metals from the 
point of view of anomalous transport theory. 

\subsection{Weyl semi-metals}

\begin{figure}[ht]
\centering
\includegraphics[width=.8\textwidth]{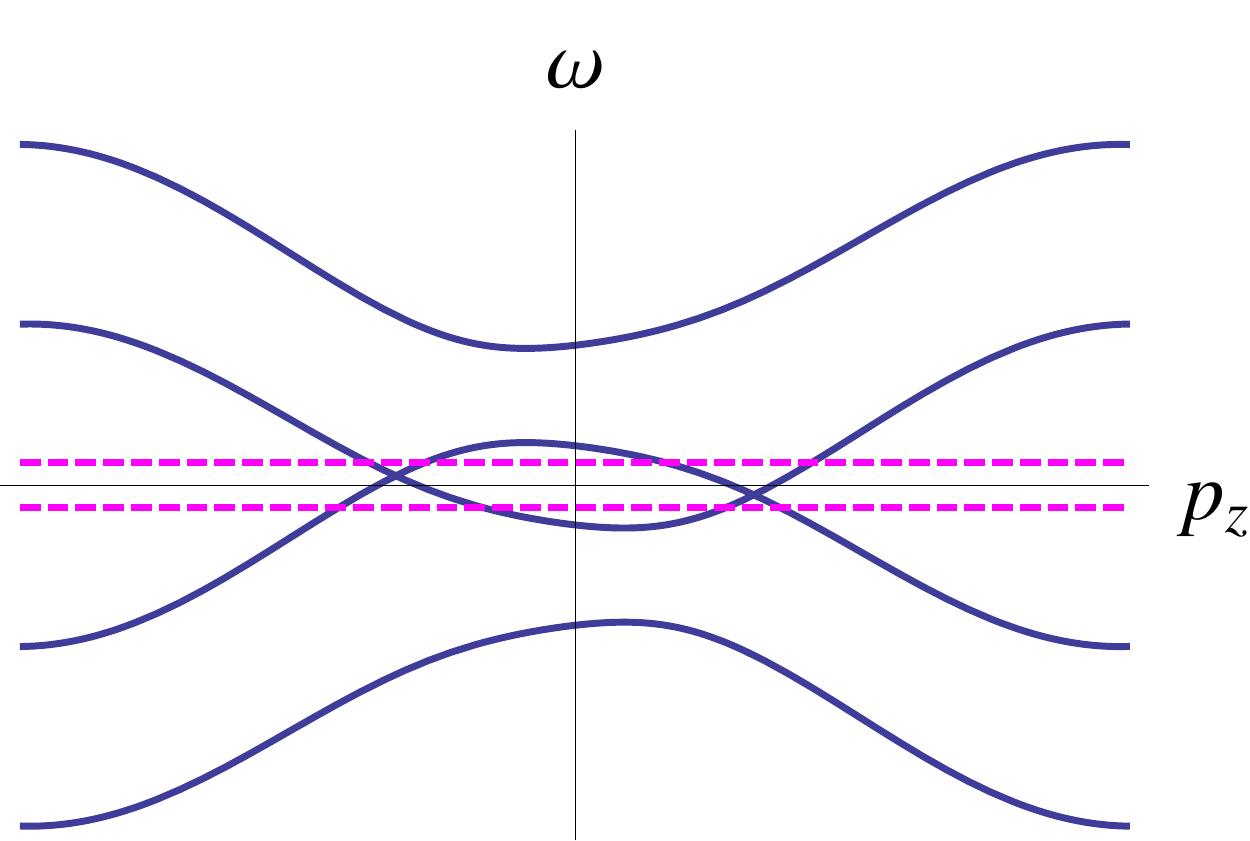}
\caption{The figure sketches the band structure of a Weyl semimetal derived from 
a tightbinding model. The energy bands are periodic along the Brillouin zone of 
which a section
with $p_x=p_y=0$ is shown. If the Fermi energy lies near the band crossing 
points the effective low energy electron dynamics is well described by Weyl 
equations of opposite
chirality. This makes the theory of anomalous transport applicable.}
\label{fig:wsmbandstructure}
\end{figure}

\begin{figure}[ht]
\centering
\includegraphics[width=.7\textwidth]{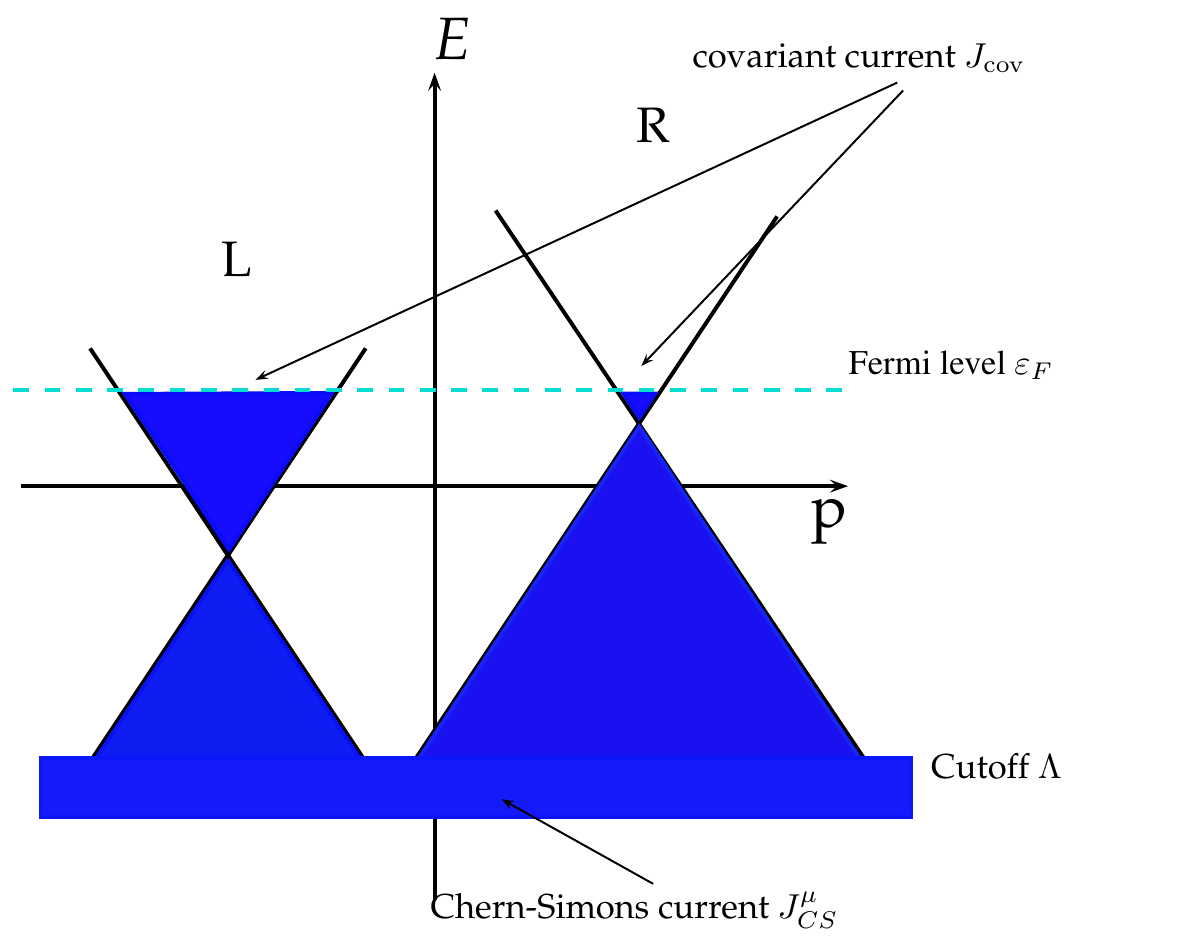}
\caption{Local (in the Brillouin zone) model of the electronic band structure of 
a Weyl semimetal. Inter valley scattering leads to equilibration of the Fermi 
levels of left
and right-handed Weylcones. The local model has to be supplemented by a charge 
preserving cutoff. The covariant current is the current that stems form 
quasiparticles excitations
near the Fermi surface. The total electric current has another component given 
by the Bardeen-Zumino polynomial or Chern-Simons current generated at the cutoff 
and 
guaranteeing charge conservation. From the point of view of the quantum field 
theoretical model the chemical potentials have to be measured with respect to 
the tips of 
the Weylcones. So in equilibrium there is a non-vanishing axial chemical 
potential $\mu  _5=b_0$ als long as the Weylcones are shifted in energy. This shift 
in energy is given
by the temporal component $b_0$ in the local model (\ref{eq:WSMmodelsimple}). On 
the other hand $b_0$ acts precisely as temporal component of an axial vector 
field
$A_0^5 = b_0=\mu_5$. }
\label{fig:wsmmodel}
\end{figure}

Weyl semi-metals are materials in which two bands cross at isolated points in 
the Brillouin zone. 
In figure \ref{fig:wsmbandstructure} we sketch a typical situation arsing from a 
tightbinding model of a Weyl semi-metal. 
If the Fermi level is at or near the band crossing point the effective 
Hamiltonian
near such a crossing point is
\begin{equation}\label{eq:Hwsmeff}
 H_\mathrm{eff} =\pm  \vec\sigma \vec k
\end{equation}
where $\vec k=(k_x,k_y,k_z)$ measures the momentum relative to the position of 
the band crossing point. This are the Hamiltonians for a Weyl fermions of
either left- or right-handed chirality.
For simplicity we have assumed rotational symmetry around the 
band crossing point. The Nielson-Ninomiya theorem \cite{Nielsen:1981hk} makes the  
band crossing points to come always in pairs with opposite choices of signs in 
(\ref{eq:Hwsmeff}), i.e. with opposite chiralities. 
In general there might be many more band crossing points but clearly the 
simplest situation is one with only two. The effective low energy degrees of 
freedom are then a left handed
and a right Weyl fermion. A more field theoretic way of describing this is the 
Dirac equation
\begin{equation}\label{eq:WSMmodelsimple}
  \left(i \slashed{\partial}  - \gamma_5  \slashed{b} \right) \Psi = 0\,.
\end{equation}
The four vector $b_\mu  $ shifts the momenta and frequencies of left- and 
right-handed fermions
\begin{equation}
 \omega_{L,R} = \pm b_0 \pm \sqrt{( \vec{p} - \vec{b})^2}\,.
\end{equation}
For a start this will serve as  model for the physics of Weyl semi-metals up to 
one more ingredient: in a crystal the brilloin zone is bounded and periodic. 
This has two effects we need to account for if we want to use the model (\ref{eq:WSMmodelsimple}): 
one is that the linear dispersion can not be extended to arbitrary high 
energies and related to it chiral symmetry  is not exact even at tree level. 
The fact that the bands are really periodic means that there is no cutoff at 
infinite energy. States can only move along the band in a periodic 
fashion\footnote{As electrons indeed do 
in very pure samples upon switching on an electric fields. The produced current 
shows oscillating behavior, the Bloch-oscillations.} but can never disappear. 
For our purpose this implies we have to 
supplement our field theoretical model with cutoff preserving the total number 
of states. More precisely an anomaly in the total charge current stemming from 
both, left-handed and
right-handed Weyl cones is not allowed. We know already that in the quantum 
field theoretical description this singles out the conserved consistent current 
as the unique candidate for the electric current. 
Furthermore we need to take into account 
that the chiral symmetry is only an accidental one. Even without anomaly it can 
not be preserved
for arbitrary momenta and energies along the band dispersion. Thus 
the description in terms of Weyl fermions is an approximate one and there is 
always a non-vanishing
amplitude for an electron to scatter out of the left-handed Weyl cones and into 
the the right-handed Weylcone. This process is known as inter valley scattering, 
it leads to an effective
decay time for axial charge. On the other hand scattering processes that leave 
the electron within its Weylcone are called intra valley scattering. It also 
usual to assume that
the scattering rate for inter valley scattering is much smaller than the 
scattering rate for intra valley scattering. Inter valley scattering will lead 
always to equilibration of the 
Fermi surfaces of left-handed and right-handed Weyl cones. Therefore the 
equilibrium situation can be sketched in figure \ref{fig:wsmmodel}. 

\subsection{Applying anomalous transport theory to Weyl semi-metals}

\subsubsection{CME:}
While the actual Fermi energies are equal for both Weylcones in the local Dirac 
model of (\ref{eq:WSMmodelsimple}) we have to assign different chemical 
potentials
to them. Intuitively the chemical potential measures the size of the Fermi 
surface and it is clear that as long as the parameter $b_0$ is different from 
zero the
size of the Fermi surfaces of left- and right-handed Weyl fermions will be 
different in equilibrium. This means the equilibrium state is described by 
a non-vanishing
axial chemical potential $\mu  _5 = b_0$. On the other hand this parameter is 
nothing else than the temporal component of an axial vector field $A_0^5 = b_0$. 
Now we can compute what happens if we switch on a magnetic field, the current 
due to the CME has be calculated as
\begin{equation}
 \vec{\mathcal{J}} = \frac{\mu  _5 - A_0^5}{2\pi^2} \vec B = 0\,.
\end{equation}
So there is no chiral magnetic effect in equilibrium.  This is  in 
accordance with explicit numerical simulations of the CME and more generally
with the so-called Bloch theorem 
\cite{Buividovich:2013hza,FranzVazifeh,Yamamoto:2015fxa} that forbids a 
non-vanishing equilibrium expectation value for a precisely conserved current. 
Instead of insisting on a description in terms
of relativistic Weyl fermions one might also stick to the convention that 
energies should be measure with respect to a common reference for both Weyl 
cones. Then
there is simply no difference in the energy one needs add an electron to the 
left-handed or right handed Fermi surface. In this sense there is no true axial 
chemical potential \cite{Basar:2013iaa,Zhou:2012ix}. 

In order to activate the chiral magnetic effect there are in principle two 
possibilities. One can either try to manipulate the relative occupation numbers 
in the
Weyl cones or one can try to change the separation in energy of the tips of the 
Weycones. Both options should give a CME signal.
Let us explore the first one. The key is to change the equilibrium state $\mu  
_5=b_0$. A conventional way of doing this is by using the axial anomaly itself.
We switch on parallel electric and magnetic fields. The change in the rate of 
axial charge is then 
\cite{Nielsen:1983rb,Son:2012bg}
\begin{equation}
 \dot{\rho}_5 = \frac{1}{2\pi^2} \vec E \vec B - \frac{1}{\tau_5} (\rho_5 - 
\rho_5^{(0)})\,.
\end{equation}
Here we have also  included the axial charge relaxation due to inter-valley 
scattering. The equilibrium axial charge density can be computed from the 
equation of state $\rho_5 = (\partial F(\mu_5,\mu,T) )/(\partial\mu_5)$ by 
inserting $\mu_5=b_0$. 
We define the axial susceptibility as $\chi_5 = \partial \rho_5/\partial \mu  
_5$ and express the axial charge via the axial chemical potential
as $\delta \rho_5 = \chi_5 \delta\mu  _5$.  In parallel electric and magnetic 
fields a stationary non-equilibrium state will be reached with
\begin{equation}
 \delta\mu  _5 = \frac{\tau_5}{\chi_5} \frac{\vec E \vec B} {2\pi^2}\,.
\end{equation}
The induced current is then the sum of ohmic and chiral magnetic current
\begin{equation}
 \mathcal{J}_i = \sigma E_i + \frac{\delta \mu  _5}{2\pi^2} B_i = \left( \sigma 
\delta_{ij} + \frac{\tau_c}{\chi_5} \frac{ B_i B_j }{4\pi^4} \right) E_j\,.
\end{equation}
where $\sigma$ is the usual Ohmic conductivity in absence of the magnetic field.
If the axial symmetry were not explicitly broken the induced conductivity would
be infinite! This reflects the non-dissipative character of the chiral magnetic 
current\footnote{In a hydrodynamic approach it has been shown that there are
additional terms in the case of finite axial chemical potential \cite{Landsteiner:2014vua}.}. 
So the chiral magnetic effect manifests itself as an enhancement of the electric 
conductivity along the magnetic field with a characteristic quadratic dependence
on the magnetic field strength\footnote{Large magnetic fields 
can induce anisotropies in $\sigma$ and they can also change the values
of the relaxation times and the axial chemical potential. For example for large 
magnetic fields the electrons will be mostly in the lowest Landau level and
then the axial susceptibility will be itself proportional to $B$.} 
This is the anomaly induced negative magnetoresistivity along the magnetic 
field. Signatures compatible with it have been measured in a variety
of experiments recently \cite{Li:2014bha,Zhang:2016ufu, 
He,Shekhar:2015rqa,Zhang:2015lya,Kim:2013dia,Ong1,Huang:2015eia,Yang:2015uka,Ong2,Felser}.

Another option is to change the energy levels of the band touching points. Using 
a tightbinding model it has recently been shown that this can be achieved by 
applying strain \cite{Cortijo:2016wnf}. In this case the current will not be stationary
but
decay over a time given by the axial charge relaxation rate of around 
$10^{-9}$ seconds.

\subsubsection{Anomalous Hall Effect}
Another effect present in in our simple model is the quantum anomalous Hall 
effect \cite{Haldane:2004zz, Grushin:2012mt,Goswami:2012db}. A current perpendicular to an 
applied electric field.
 In our quantum field theoretical setup it is exclusively due to the Bardeen-Zumino 
polynomial and takes the form
\begin{equation}\label{eq:wsmahe}
 \vec{\mathcal{J}} = \frac{1}{2\pi^2} \vec E \times \vec b\,.
\end{equation}
It follows from the Chern-Simons current in (\ref{eq:covcurrent}) by identifying 
$\vec A_5 = \vec b$. 
We note that the parameter $\vec b$ breaks time reversal symmetry. In time 
reversal invariant Weyl semi-metals  there are
several pairs of Weyl nodes such  that the vector sum $\sum \vec b_i =0$ and the 
axial anomaly induced Hall effect is absent.
The Bardeen-Zumino polynomial also implies an induced charge in magnetic field
of the form
\begin{equation}
\mathcal{J}^0 = \frac{1}{2\pi^2} \vec b . \vec B\,.
\end{equation}
a three dimensional generalization of what is known as Streda formula in condensed 
matter theory \cite{Streda}.

\subsubsection{Fermi Arcs}

From the identification $\vec A_5 = \vec b$ we can learn a lot more. First we 
note that  a Weyl fermion can not be given a mass in a gauge invariant
way. This is what makes the Weyl nodes stable in a Weyl semimetal. On the other 
hand we know that outside any finite piece of material there are no low
energy fermions, the vacuum is a (trivially) gapped state in which the electrons 
are massive. The only way to make the Weyl fermions massive is to bring
them together at the same point in momentum space and switching on a Dirac
mass term. 
This means that the vector $\vec b$ must go to zero as one crosses the edges of the 
material. On the edge of  a Weyl semimetal there
is necessarily a gradient of $\vec b$. This is nothing else but an axial 
magnetic field $\vec B_5 = \nabla \times \vec A_5$ \cite{Chernodub:2013kya}. An axial 
magnetic field acts like
a usual magnetic field but with opposite signs for Weyl fermions of opposite 
chirality. Lets make a simple model in which there is a sharp edge at say $x=0$
and $\vec b = b \Theta(y) \hat e_z$. Thus there is a strong axial magnetic field 
localized at the edge $\vec B_5 = b \delta(y) \hat e_x$. 
The Weyl fermions will have zero modes localized at the axial magnetic field 
lines (the states in the lowest Landau level). In a usual magnetic field these 
zero modes are $1+1$ dimensional
chiral fermions of opposite chirality but in the axial magnetic field the zero 
modes with have the same chirality. The local density of states is the one
of the lowest Landau level $B_5/(2\pi)$ for each  Weyl cone. So in total we 
predict localized chiral edge states with linear dispersion only along the
axial magnetic field lines and degeneracy $b/\pi$! These are the famous 
Fermi-arcs \cite{WSM1}. In actual materials Fermi arcs turn out not 
to be just straight lines and this has been connected to boundary conditions for the 
Weyl fermions in \cite{Hashimoto:2016kxm}. It would be interesting to 
see how the simple picture of the axial magnetic field can be modified to catch 
these subtleties.

We can also apply anomalous transport formulas to the physics of Fermi-arcs. Let 
us assume we have a doped Weyl semimetal. Then we can
employ a version of the chiral magnetic effect for axial magnetic fields
\begin{equation}\label{eq:wsmameJ}
 \vec{\mathcal{J}} = \frac{\mu  }{2\pi^2} \vec B_5\,.
\end{equation}
This is an edge current. We can also derive the anomalous Hall effect from this. 
We assume that there are two edges at say $y=0$ and at $y=L$ and in between
a WSM with $b_z>0$. In this situation there will be no net current since the 
edge currents at $y=0$ and at $y=L$ cancel each other exactly. Let us now 
inject additional electrons at the edge at $x=0$. This will rise the local 
chemical potential there and therefore increase the local edge current. Now there 
is net current the form
\begin{equation}
 \mathcal{I}_x = \int dy\, \mathcal{J}_x = \frac{\Delta_y \mu  }{2\pi^2} b_z
\end{equation}
So we get a net current perpendicular to the gradient of the chemical potential. 
This is very similar to (\ref{eq:wsmahe}) except that now the current is
localized at the edges whereas in  (\ref{eq:wsmahe}) it is better thought of as 
a bulk current. As usual for the quantum Hall effect it appears either as edge 
current
or as bulk current depending on whether one pumps charge from one edge to the 
other via a bulk current generated by an external electric field or one directly 
injects fermions at the edges. 

\subsubsection{Thermal Hall Effect}

Let us study now the energy transport instead of the charge transport. One of 
the most interesting aspects is that thermal transport is
sensitive to the gravitational anomaly. In particular in an axial magnetic field 
we have
\begin{equation}\label{eq:wsmameT}
 \vec{J}_\varepsilon =\left( \frac{\mu  ^2}{4\pi^2} + \frac{T^2}{12} \right) 
\vec{B}_5 \,.
\end{equation}
since in Weyl semi-metals there necessarily exists a strong axial magnetic field 
at the edge we can use this to derive the thermal Hall effect.
As before suppose that there are two edges at $y=0$ and at $y=L$. In equilibrium 
the net energy current vanishes since the currents at opposite
edges cancel each other exactly. Let us heat up the edge at $y=L$ to a 
temperature $T+\Delta T$. Now there is a net energy current
along the $x$ direction 
(keeping the chemical potential fixed)
\begin{equation}
 J_{\epsilon,x} = \frac{1}{6} T \Delta_y T b_z 
\end{equation}
from which we can infer the thermal Hall conductivity
\begin{equation}
 \kappa_{T,Hall} = \frac{T}{6} |b|
\end{equation}
Contrary to the Hall effect this is a pure edge current. The bulk current in the 
Hall effect was related to the Bardeen-Zumino polynomial that converts the
non-conserved covariant current into the conserved consistent current. There is 
an analogue Bardeen-Zumino current for the consistent energy momentum
tensor.  It is (just as the gravitational anomaly itself) of higher order 
in derivatives. Thus there is no thermal Hall current in the bulk as a response 
to a gradient in the temperature.

\subsubsection{Axial negative magnetoresistivity}
The covariant from of the anomaly also suggests that one can induce negative 
magnetoresistive via an axial magnetic field since
\begin{equation}\label{eq:wsmcovanomaly}
 \dot{\rho} = \frac{ \vec E \vec B_5}{2\pi^2}\,,
\end{equation}
combined with the axial magnetic effect (\ref{eq:wsmameJ}) should give a large 
enhancement of the conductivity. At first glance there
is seemingly a problem. In the case of the negative magnetoresistivity induced by 
parallel usual electric and magnetic fields the axial
charge decays also due to tree level non-conservation. This makes the induced 
conductivity finite. In the case of the electric charge
this is impossible, because of gauge invariance the electric charge can 
never decay. This seems to give the unphysical result
of an infinite conductivity in parallel electric and axial magnetic fields. 
The resolution comes from the nature of the axial field. As we noted
it can exist only inside the Weyl semimetal and has to vanish necessarily in the 
vacuum outside the material. The net total charge
induced by (\ref{eq:wsmcovanomaly}) is 
\begin{equation}\label{eq:totalB5zero}
 \frac{d}{dt} Q = \int_\Omega d^3x \dot \rho = \int_\Omega d^x \epsilon_{ijk} 
\partial_j A^5_k E_i = \oint_{\partial\Omega} dS_j \epsilon_{ijk} A^5_k E_i 
=0\,,
\end{equation}
where $\Omega$ denotes a region of space that contains the Weyl semimetal. This 
region can always be taken much larger than the space region occupied by
the material. Since the axial vector vanishes on $\partial\Omega$ no net charge can be 
induced. Every region with a positive covariant anomaly must necessarily be
offset by a region with equal amount of negative covariant anomaly. From the 
point of view of the consistent anomaly everything that
is happening is that a current is created transporting charge from one region to 
another. Since now the charge distribution is inhomogeneous we need
to take diffusion into account
\begin{align}
 \dot\rho + \vec\nabla \vec J = \frac{\vec{E}.\vec{B}_5}{2\pi^2} \, \\
 \vec J = - \sigma \vec\nabla \mu   + \frac{\mu  }{2\pi^2} \vec{B}_5 \,.
\end{align}
Or from the point of view of the consistent current where the covariant anomaly 
is interpreted as inflow via the anomalous Hall effect 
\begin{align}
 \dot\rho + \vec\nabla \vec{\mathcal{J}} =0 \, \\
 \vec{\mathcal{J}} = - \sigma \vec\nabla \mu   + \frac{\mu  }{2\pi^2} \vec{B}_5 
- \frac{1}{2\pi^2} \vec b \times\vec E \,.
\end{align}
The local growth of charge due to Hall current inflow is counterbalanced 
eventually due to diffusive charge outflow such that in a stationary situation 
$\dot \rho=0$.
We also note that these considerations also apply to the case where $\vec{B}_5$ 
is created in the bulk of the material via strain \cite{Cortijo:2016yph}.
Negative axial magnetoresistivity has recently also been discussed in 
\cite{Pikulin:2016wfj,axialmagneticAdolfo}.

Let us work out a simple example. We describe the Weyl semi-metal as the region 
with $\vec{b} = b \hat e_z [\Theta(y) - \Theta(y-L)$.
At the edges there is strong axial magnetic field $\vec{B}_5 = B_5 \hat e_x$ 
with $B_5 = b [\delta(y) - \delta(y-L)]$.
In and electric field parallel to the axial magnetic field the stationary 
solution is
\begin{equation}
 \sigma \partial_y^2 \mu  = - \frac{B_5 E}{2\pi^2} \,.
\end{equation}
which is solved by
\begin{equation}
 \mu  = - \frac{ b E}{\sigma 2 \pi^2} [ y \Theta(y) - (y-L) \Theta(y-L) 
-\frac{L}{2}]\,.
\end{equation}
Here $\Delta \mu  = \mu  (0) - \mu  (L) = \frac{L b E}{\sigma 2 \pi^2}$ is the 
Hall voltage. 
The total current in the along the electric field is 
\begin{equation}
 \mathcal{J}_x = [\sigma +  \frac{ L b^2}{\sigma 8 \pi^4}(  \delta(y)  + 
\delta(y-L) ) ] E\,.
\end{equation}
The conductivity is strongly enhanced at the edges due to the axial magnetic 
effect. For fixed width $L$ the enhancement is stronger as the bulk conductivity 
gets smaller. In the limit of vanishing bulk conductivity the edges are perfect 
conductors. This is the limit in which only the edge states contribute to the 
conductivity.
They are chiral fermions and indeed have formally infinite conductivity. 
The true physical conductivity will stay finite of because of the finite range in
energy of the Weyl cones. Once the local Fermi level exceeds the range of linear disperison
the effect of the anomaly stops.
On the other hand for finite bulk conductivity $\sigma$ the charge of the edge states
can diffuse into the bulk and this is what makes the anomalous enhancement 
finite\footnote{A similar conclusion has been reached in a more microscopic model
in \cite{Gorbar:2016aov}}.

\subsection{Chiral collective waves}
Another direct consequence of the chiral magnetic effect is the the so called 
chiral magnetic wave. We combine the chiral magnetic and the chiral separation
effect with current conservation (assuming absence of an electric field parallel 
to the magnetic field. Assuming that vector and chiral charges are
linearly related to the chemical potentials and a spatially homogeneous 
magnetic field we find
\begin{align}
 \chi_5 \dot\mu  _5 + \frac{\vec B}{2\pi^2} \vec\nabla\mu  &=0 \,\\
 \chi \dot\mu  + \frac{\vec B}{2\pi^2} \vec\nabla\mu  _5 &=0 \, .
\end{align}
Fourier transforming in space and time and setting the determinant to zero gives 
the mode equation
\begin{equation}\label{eq:cmv}
 \omega^2  4 \pi^4 \chi \chi_5 - (\vec{k}\vec{ B})^2 =0
\end{equation}
This is the chiral magnetic wave. It can be understood as propagating 
oscillation between axial and vector charge. In the large magnetic 
field limit
it follows from (\ref{eq:fwithB}) that its velocity approaches the speed of 
light. 
The chiral magnetic wave should be observable as a collective mode of the 
electron fluid in Weyl semi-metals. It also has  an important
application in the physics of the quark gluon plasma, where it leads to a 
quadrupole moment in the charge distribution of the final hadronic state
in non central heavy ion collisions. This signal has been theoretically predicted 
in \cite{Gorbar:2011ya,Kharzeev:2010gd,Burnier:2011bf} and observations in experiment 
compatible with it are reported
in \cite{Adam:2015vje,Adamczyk:2015eqo}.

Chiral hydrodynamics has been used to find many more interesting collective modes 
such as appearance of non-linear Burgers solitary waves due to the chiral 
vortical effect \cite{Basar:2013iaa},
a chiral vortical wave \cite{Jiang:2015cva} at non-zero chemical potentials, a 
chiral heat wave \cite{Chernodub:2015gxa,Kalaydzhyan:2016dyr}, chiral Alfven 
waves \cite{Yamamoto:2015ria, Abbasi:2015saa}. 

\section{The holographic Weyl semimetal}

So far we have worked with a very simple field theory model for a (time-reversal 
breaking) Weyl semimetal based on the theory of anomalous transport.
Holography has been of highest importance to gain understanding of anomaly 
induced transport and even to unravel some unexpected relations such
as the one between the temperature dependence of the chiral vortical effect and 
the gravitational anomaly. This motivated work on a holographic
model of Weyl semimetal and the relation between anomalies and 
transport phenomena in this model \cite{Landsteiner:2015lsa, 
Landsteiner:2015pdh,Landsteiner:2016stv}.

The important feature is that the Weyl physics arises in the 
infrared region. This is unusual from the point of view of high
energy physics where massless fermions are usually thought of as being a good 
approximation for high energy processes. A simple field theoretic model
of how effective low energy Weyl physics can arise is the Lorentz breaking 
massive Dirac equation
\begin{equation}\label{eq:lorentzbreaking}
 (i \slashed{\partial}   - M + \gamma_5 \slashed{b} ) \Psi =0
\end{equation}
This model has been invented to investigate the consequences of Lorentz 
symmetry breaking in particle physics  \cite{Colladay:1996iz} and has 
been unsed as a model for 
the physics of Weyl semi-metals in \cite{Grushin:2012mt}.

\begin{figure}\begin{center}
 \includegraphics[width=.45\textwidth]{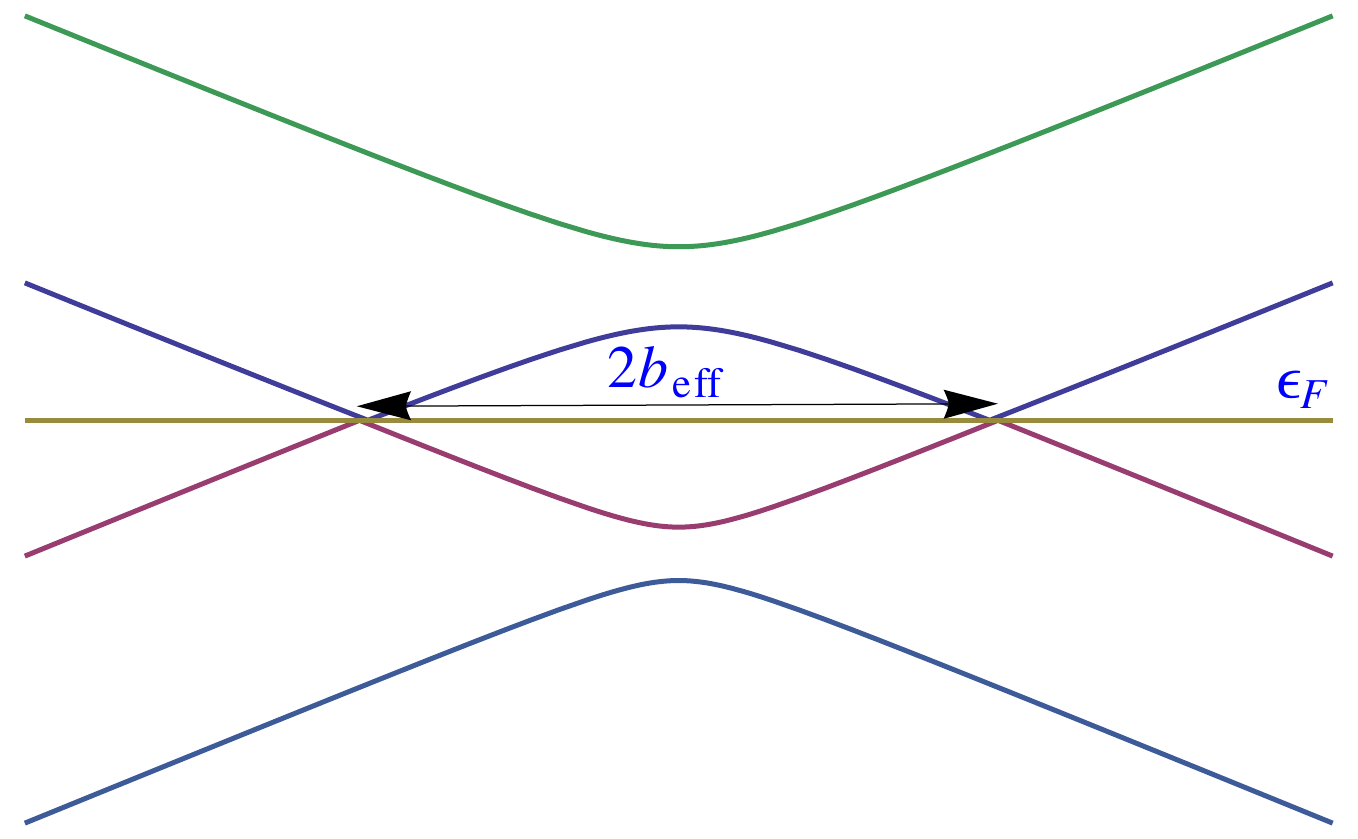}
 \includegraphics[width=.45\textwidth]{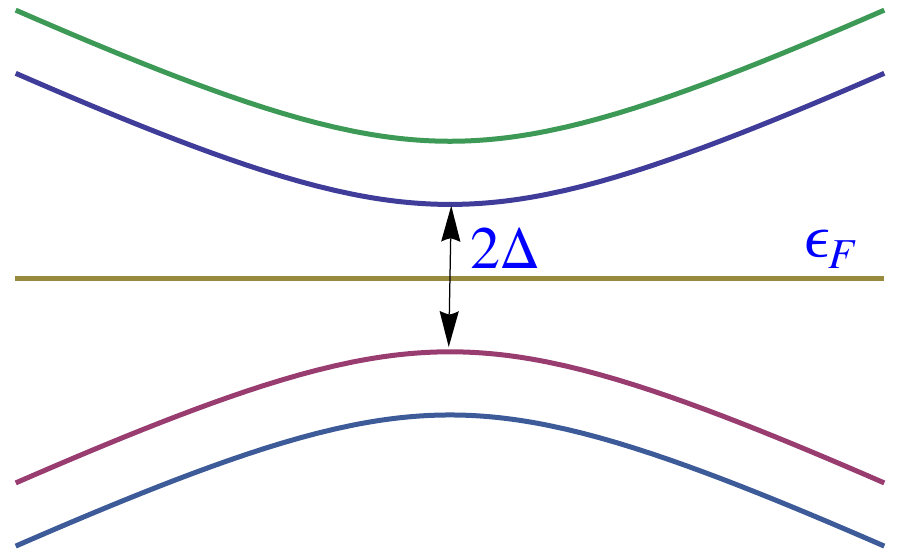}
 \caption{\small Left panel: For $b^2>M^2$ there are two Weyl nodes in the 
spectrum. They are separated by the distance $2\sqrt{b^2-M^2}$ in momentum 
space.
Right panel: For $b^2<M^2$ the system is gapped with gap $2\Delta = 
2\sqrt{M^2-b^2}$.}
\label{fig:topphase}
\end{center}
\end{figure}
We will concentrate on the case in which the four vector $b_\mu  $ is purely space 
like and without loss of generality we take it to point int the $z$ direction.
At high energies the mass term is irrelevant and the theory is basically the 
same as (\ref{eq:WSMmodelsimple}). If we are interested
in the behaviour at low energies there is some interesting non-trivial phase 
transition. It turns out that as long as $|M|<|\vec{b} |$ the low energy theory
is not gapped! Rather the low energy theory is given  again by   
(\ref{eq:WSMmodelsimple}) but with an effective low energy parameter
$b_\mathrm{eff} = \sqrt{b^2-M^2}$. So the low energy theory is one of massless 
Weyl fermions despite the fact that there is  a mass parameter in
the fundamental Lagrangian. On the other hand for $|M|>|b|$ the theory is gapped 
with a mass gap of $\Delta = \sqrt{M^2-b^2}$. 
There is a quantum phase transition at $M=b$. In fact it is a topological 
phase transition since the topology of momentum space changes from a situation
with band crossing points to one with a gap (see figure \ref{fig:topphase}). 
What is the signature of this topological phase transition? We can argue that it 
should be the anomalous
Hall effect. Since the Hall effect is the response to static and homogeneous 
fields it is a IR property and it should be governed by the properties of
the IR theory. We already know that a Dirac fermion with a constant axial vector 
$\vec A^5 = \vec b$ features an anomalous Hall effect (\ref{eq:wsmahe}). 
Since In the IR theory the relevant parameter is $b_\mathrm{eff}$ so we expect 
that there is an anomalous Hall conductivity of the form
\begin{equation}
 \sigma_{xy} = \frac{\sqrt{b^2-M^2}}{2\pi^2} \Theta(|b|-|M|) \,.
\end{equation} 

The calculation of this Hall conductivity in field theory (even the free one) is 
not easy and subject to all the renormalizaiton ambiguities we discussed already
in the case of the triangle diagrams \cite{Jackiw:1999qq}. Instead of analyzing the field theory any 
further we will take it as motivation to write down
a holographic model that also has a topological quantum phase transition between 
a topological state with non-zero Hall conductivity and a topologically
trivial state with vanishing Hall conductivity. 

Now we want to construct a holographic model based on the dictionary in table 
\ref{tab:dictionary}. We want to implement the particular symmetries and
their breaking patterns based on the free fermion model 
(\ref{eq:lorentzbreaking}). Most basically the theory lives in flat Minkowski 
space. Lorentz symmetry is
only broken by the vector $b_\mu  $ which we understand as background of an 
axial gauge field. There are also two $U(1)$ symmetries, a vector like one 
representing
the electric charge conservation and an axial one that is broken at tree level 
by a mass term. The mass operator $\bar \Psi \Psi$ is rotated into $\bar \Psi 
\gamma_5 \Psi$
under the action of the axial $U(1)$ symmetry. The field content of our 
holographic model has to have a metric to represent energy and momentum 
conservation
related to Lorentz symmetry, a completely conserved vector gauge field $V_\mu  
$, an axial gauge field $A_M$, a complex scalar field whose real part 
corresponds to the mass operator. The scalar is also charged under the axial 
$U(1)$. The axial anomaly has three parts, a pure axial cubed anomaly, a
mixed axial vector anomaly and a mixed axial gravitational anomaly and is 
represented in holography via Chern-Simons terms. This motivates the action
\begin{align}
 \label{eq:holomodel}
  &&S=\int d^5x 
\sqrt{-g}\bigg[\frac{1}{2\kappa^2}\Big(R+\frac{12}{L^2}\Big)-\frac{1}{4e^2}
\mathcal{F}^2-\frac{1}{4e^2}F^2 
  \nonumber\\
 &&~~  -(D_M\Phi)^*(D^M\Phi)-V(\Phi)  \\
 &&+\epsilon^{MNPQR}A_M \bigg(\frac{\alpha}{3} \Big(3 \mathcal{F}_{NP}  
\mathcal{F}_{QR}+F_{NP} F_{QR}\Big)+\zeta  R^{L} _{~KNP}R^{K}_{~LQR} 
\bigg)\bigg]\,.\nonumber
\end{align}
The covariant derivative is $D_M = \partial_M + i q A_M $ since the scalar is 
charged only under the axial symmetry. 
The scalar field potential is chosen to be $V(\Phi) = m^2 |\Phi|^2 + 
\frac{\lambda}{2} |\Phi|^4$. The mass determines the dimension of the operator 
dual to $\Phi$ and we chose it to be
$m^2L^2 = -3$. Here $L$ is the scale of the AdS space. In the following 
we set $2\kappa^2=e^2=L=1$. The operator dual to $\Phi$ has therefore dimension
three just as the mass term in the Dirac equation (\ref{eq:lorentzbreaking}).
The electromagnetic and axial currents are defined as
\begin{align}\label{eq:consVcur}
J^\mu  &= \lim_{r\rightarrow\infty}\sqrt{-g}\Big(F^{\mu  r}+4\alpha\epsilon^{r 
\mu  \beta\rho\sigma} A^5_{\beta} F_{\rho\sigma}  \Big) \,,\\
\label{eq:consAcur}
 J^\mu  _5 &= \lim_{r\rightarrow\infty}\sqrt{-g}\Big(F_5^{\mu  
r}+\frac{4\alpha}{3}\epsilon^{r \mu  \beta\rho\sigma} 
A^5_{\beta}F^5_{\rho\sigma}  \Big)\,.
\end{align}
We are looking for solutions that are asymptotically AdS. In addition the 
holographic analogues of the mass term and the time-reversal breaking parameters 
in (\ref{eq:lorentzbreaking}) are
introduced via the boundary conditions at $r=\infty$, 
\begin{equation}\label{eq:bcs}
 \lim_{r\rightarrow \infty}\,r\Phi = M~,~~~\lim_{r\rightarrow \infty}A_z = b\,.
\end{equation}
Our ansatz for the zero temperature solution is
\begin{eqnarray}\label{eq:ansatz}
ds^2&=&u(-dt^2+dx^2+dy^2)+\frac{dr^2}{u}+h dz^2\,,\nonumber\\
 A&=&A_z dz\,,~~~\Phi=\phi\,\,.
\end{eqnarray}
Note that due to the conformal symmetry at zero temperature only $M/b$ is a 
tunable parameter of the system.\\[1cm]

\noindent{\bf Critical solution:}
The following Lifshitz-type solution is an exact solution of the system. 
\begin{eqnarray}\label{nh-cs}
ds^2&=&u_0r^2(-dt^2+dx^2+dy^2)+\frac{dr^2}{u_0r^2}+h_0 
r^{2\beta}dz^2\,,\nonumber\\
A_z&=&r^\beta,~~~\phi=\phi_0\,.
\end{eqnarray}
 It has the anisotropic Lifshitz-type symmetry $(t,x,y,r^{-1})\to 
s(t,x,y,r^{-1})$ and $z\to s^\beta z$.
We need to introduce irrelevant deformations to flow to the UV and match the 
boundary conditions (\ref{eq:bcs}). 
We can use the scaling symmetry $z\to s z$ to set the coefficient in $A_z$  to 
be 1. There are four constants $\{u_0, h_0, \beta, \phi_0\}$ 
determined 
by the value of $\lambda$, $m$ and $q$.
To flow this geometry to asymptotic AdS in the UV, we need to consider the 
following irrelevant perturbation around the Lifshitz fix point
$
u =u_0r^2\big(1+ \delta u\,r^\alpha\big),~~
h= h_0 r^{\beta}\big(1+ \delta h\, r^\alpha\big), ~~
A_z =r^\beta \big(1+ \delta a\, r^\alpha\big),~~
\phi=\phi_0\big(1+ \delta \phi\, r^\alpha\big)
$.
Because of the scaling symmetry, only the sign of $\delta \phi$ is a free 
parameter. Numerics shows 
that only $\delta\phi=-1$ corresponds to asymptotic AdS space at the UV.  
We also fix $q=1,\lambda=1/10.$ In this case the numerical values are $(u_0, 
h_0, \beta, \phi_0,\alpha)\simeq (1.468,0.344, 0.407, 0.947,1.315)$ and 
$(\delta u, \delta h, \delta a)\simeq (0.369, -2.797, 0.137)\delta \phi$.  
Integrating towards the UV gives the value $M/b\simeq 0.744$.\\[1cm]

{\noindent{\bf Topological nontrivial phase:} }
The topologically nontrivial solution\footnote{Similar near horizon geometries were found 
in \cite{Gubser:2009cg,Basu:2009vv} in the context of holographic 
superconductors.} at in the IR is 
\begin{align}\label{nearhor-nt}
u=r^2,~~
h=r^2,~~
A_z=a_1+\frac{\pi a_1^2\phi_1^2}{16 r} e^{-\frac{2 a_1 q}{r}},~~\nonumber\\ 
\phi=\sqrt{\pi}\phi_1\Big(\frac{a_1 q}{2r}\Big)^{3/2} e^{-\frac{a_1 q}{r}}\,; 
\end{align}
$\lambda$ appears only at higher order terms and $a_1$ can be set to a numerically
convenient value. Once the solution is found it can be re-scaled to $b=1$. 

Starting from this near horizon solution we can numerically 
integrate the equations towards the UV and taking $\phi_1$ as shooting parameter.
One gets an 
AdS$_5$ to AdS$_5$ domain wall which for the chosen values of $\lambda$ and $q$ this 
exists only for $M/b<0.744$. \\[1cm]
{\noindent{\bf Topological trivial phase:}  }
The  near horizon expansion of the trivial solution is  
\begin{equation}\label{nearhor-tt}
u=\big(1+\frac{3}{8\lambda}\big)r^2,~~
h=r^2,~~
A_z=a_1 r^{\beta_1},~~\phi=\sqrt{\frac{3}{\lambda}}+\phi_1 r^{\beta_2}\,,
\end{equation}
where $(\beta_1,\beta_2)=(\sqrt{1+\frac{48q^2}{3+8\lambda}}-1, 
2\sqrt{\frac{3+20\lambda}{3+8\lambda}}-2).$
For the chosen $\lambda$ and $q$  numerically
$(\beta_1,\beta_2)=(\sqrt{\frac{259}{19}}-1, \frac{10}{\sqrt{19}}-2)$. 
We set $a_1$ to $1$ and take $\phi_1$ as the shooting parameter. Again
one finds an AdS$_5$ to AdS$_5$ domain wall.  
This type of solution only exists for $M/b>0.744$.

\begin{figure}
\begin{center}
\includegraphics[width=0.45\textwidth]{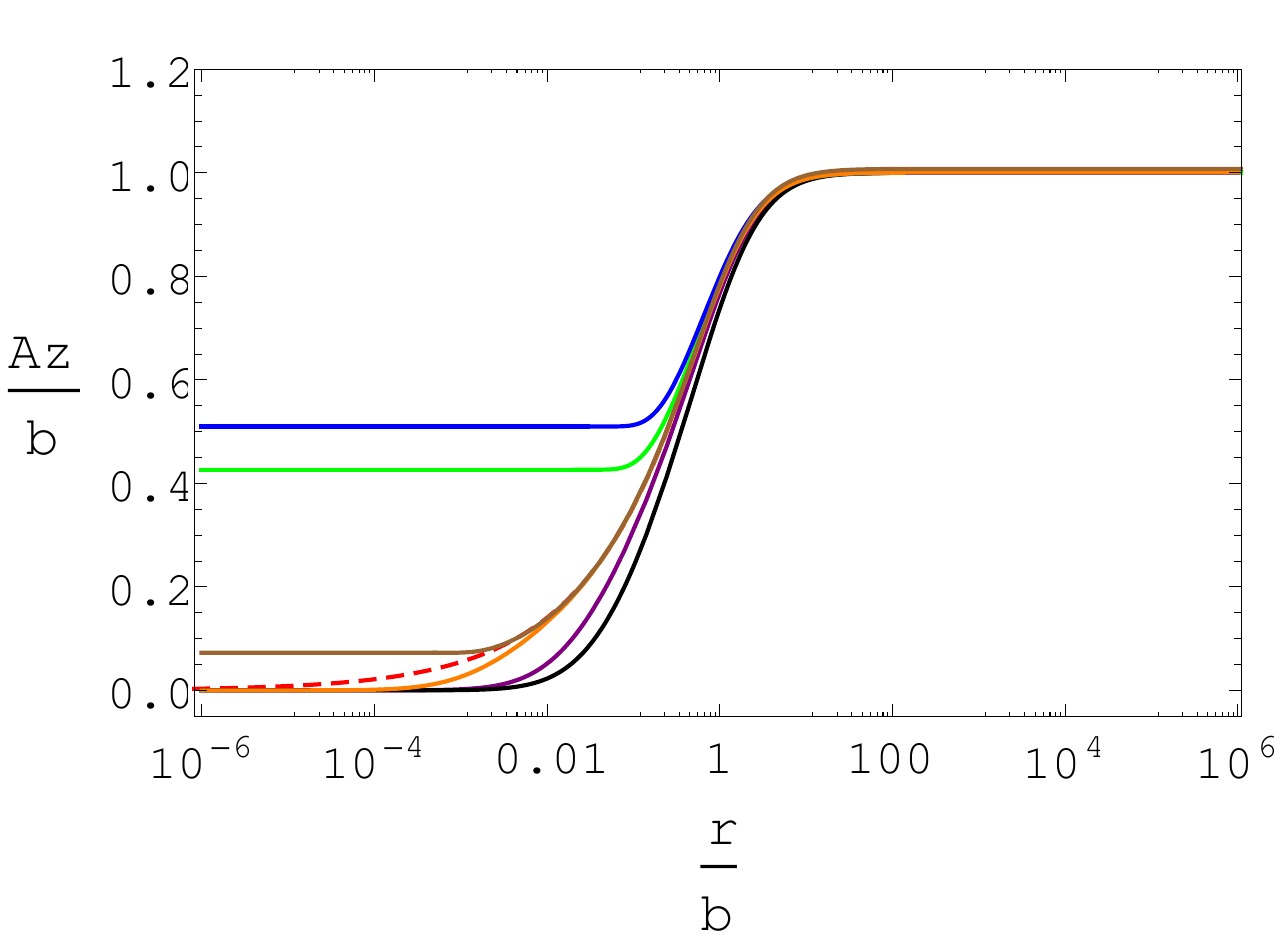}
\includegraphics[width=0.45\textwidth]{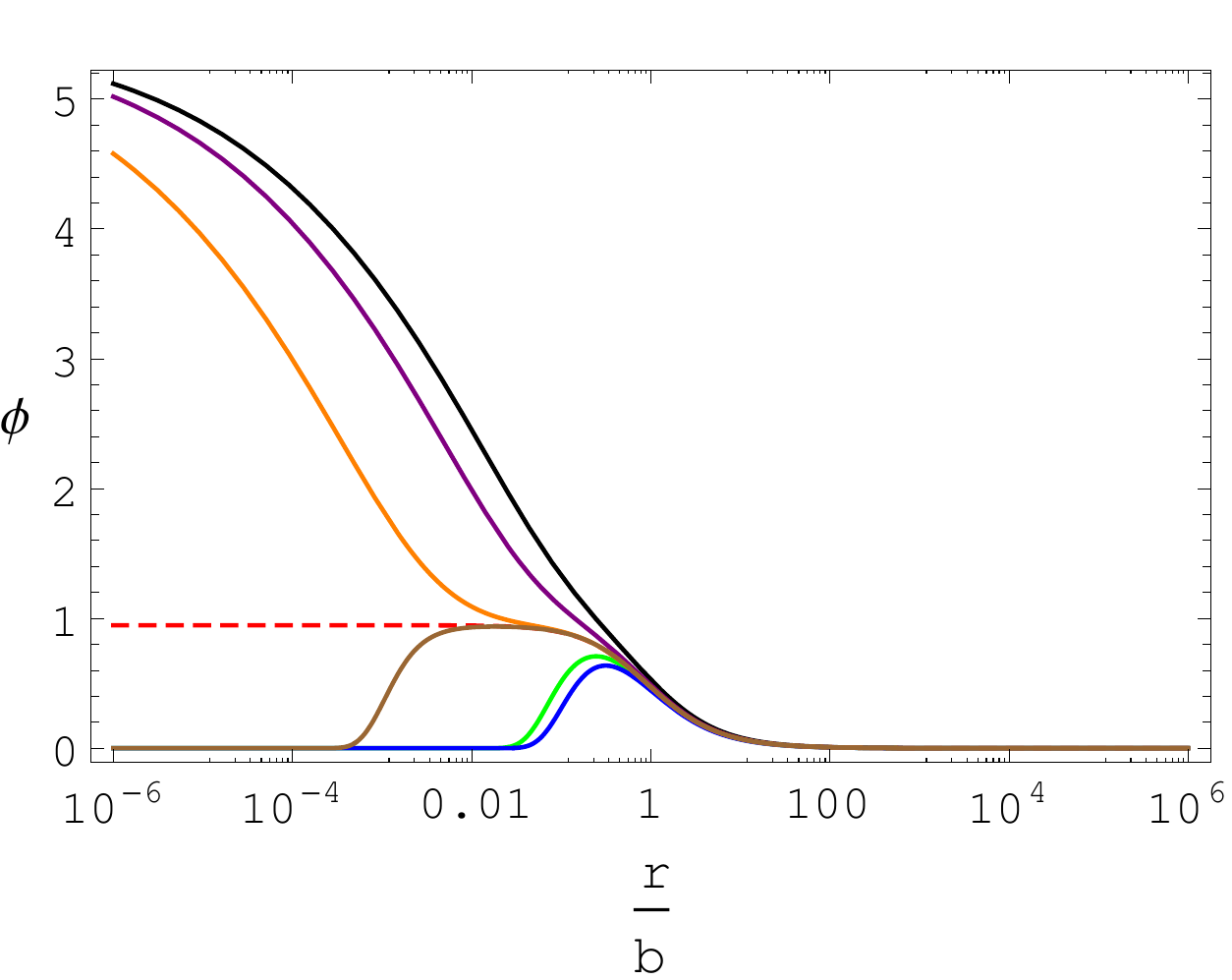}
\end{center}
\caption{\small The bulk profile of background $A_z$ and $\phi$ for $M/b=0.695$ 
(blue), $0.719$ (green), $0.743$ (brown), 
$0.744$ (red-dashed), $0.745$ (orange), $0.778$ (purple), $0.856$ (black).}
\label{fig:profile}
\end{figure}

Fig. \ref{fig:profile} shows the behavior of the scalar field and the 
gauge field for all three phases at several different values of $M/b$. For a 
given value of $M/b$ only one of the three types of solutions exists. 
The value of the gauge field on the horizon matches continuously between the
two phases whereas the value of the scalar field on the horizon jumps 
discontinuously. \\[1cm]

\noindent{\bf Finite temperature solutions:} Finite temperature 
solutions with a regular horizon can be found with the ansatz
\begin{eqnarray}
\label{ansatzforfiniteT}
ds^2&=&-udt^2+\frac{dr^2}{u}+f(dx^2+dy^2)+h dz^2\,,\nonumber\\
A&=&A_z dz\,,~~~\Phi=\phi\,,
\end{eqnarray}
imposing the conditions that at $r=r_0$ $f,h,\Phi,A_z$ are analytic and $u$ has 
simple zero. Using the scaling symmetries of AdS and the constraint from
the equations of motion at the horizon $r=r_0$ we are left with only two 
dimensionless parameters. In the UV these are mapped to $M/b$ and $T/b$. \\[1cm]

\noindent{\bf Conductivities} can now be computed with the help of Kubo 
formulas
\begin{equation}\label{eq:sigma}
 \sigma_{mn} = \lim_{\omega\rightarrow 0}\frac{1}{i\omega} \langle J_m J_n 
\rangle (\omega,\vec k =0)\,.
\end{equation}
In holography the retarded Green's functions can be obtained by studying the 
fluctuations of the gauge fields around the background with infalling boundary 
conditions at the horizon \cite{Son:2002sd}.

The anomalous Hall conductivity is the off-diagonal part of (\ref{eq:sigma}).
To compute it we need to switch on the following fluctuations 
$\delta V_x=v_x(r) e^{-i\omega t}, 
~\delta V_y=v_y(r) e^{-i\omega t}$ and define $v_\pm =v_x+i v_y$.
The equation of motion for this fluctuation is
\begin{equation}\label{eq:flucsxy}
v_\pm ''+\bigg(\frac{h'}{2h}+\frac{u'}{u}\bigg)v_\pm '+\frac{\omega^2}{u^2}v_\pm 
\pm \frac{8\omega\alpha }{u\sqrt{h}}A_z'v_\pm =0\,.
\end{equation}

\begin{figure}
\begin{center}
\includegraphics[width=0.7\textwidth]{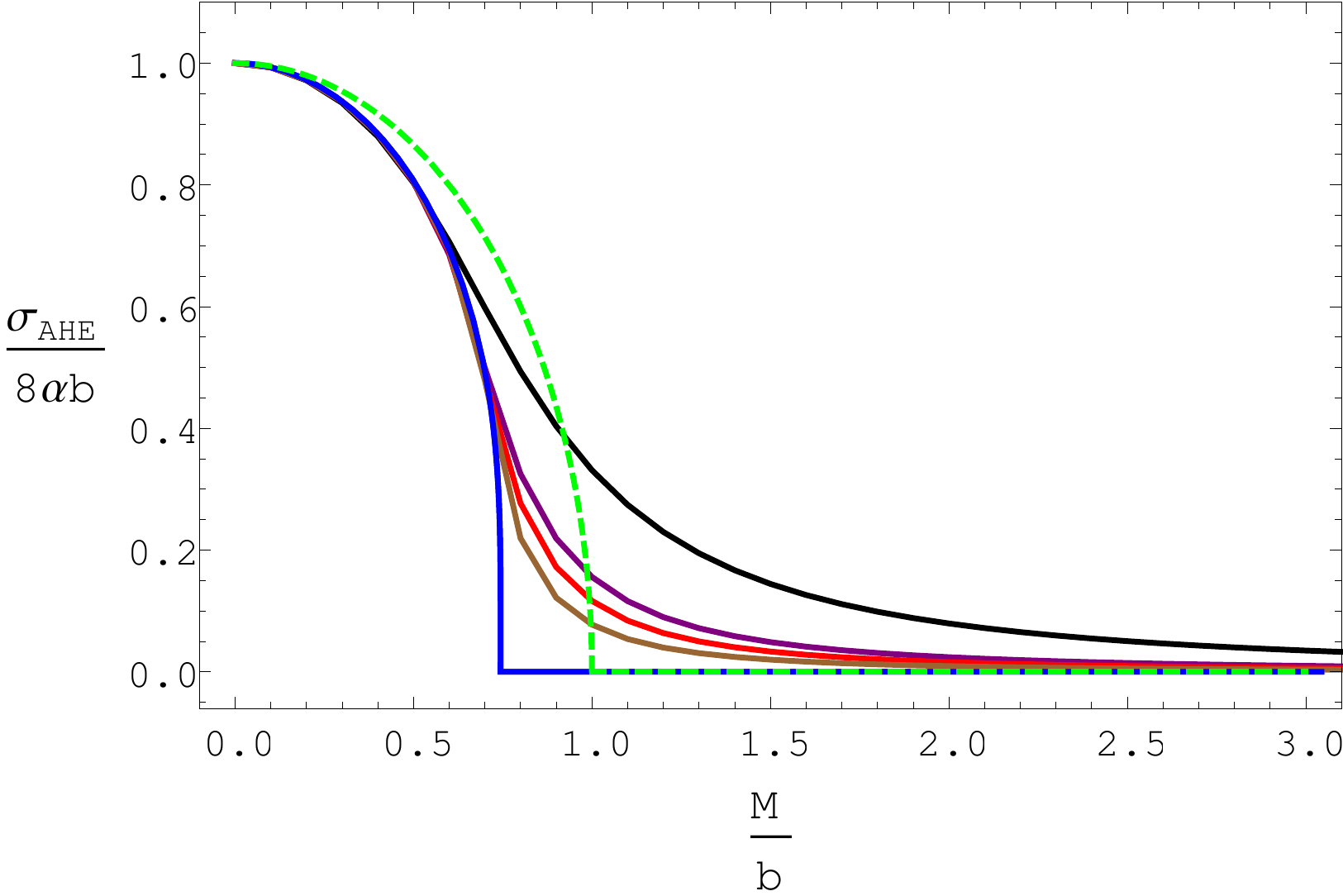}
\end{center}
\caption{\small Anomalous Hall conductivity for different temperatures. The 
solid lines correspond to our holographic model. For $T=0$ there is a
sharp but continuous phase transition at a critical value of $M/b$ (blue) which 
becomes a smooth crossover at $T>0$. We show the curves for
$T/b=0.1$ (black), $0.05$ (purple), $0.04$ (red), $0.03$ (brown). For 
comparison we also show the result for the weak coupling model as a dashed 
(green) line.
Near the transition the Hall conductivity behaves as $(\sigma_\text{AHE}/b) 
\propto ((M/b)_c-M/b)^\alpha$ with $\alpha \approx 0.211$ (to be contrasted with 
the field theory model for which $\alpha=0.5$).}
\label{fig:ahe}
\end{figure}

These are the same for the zero and finite temperature backgrounds.
To solve these equations we follow the usual near-far matching 
method \cite{Faulkner:2009wj}.
The Green's function can be read off by normalizing the fluctuation to unity at the 
boundary. The response in the current
is then given by 
\begin{equation}
G_\pm = - u\sqrt{h}  v_\pm '|_{r=\infty} \mp 8 \alpha b \omega\,.
\end{equation}
The second term stems from the Chern-Simons current in 
(\ref{eq:consVcur}), i.e. it is the contribution of the Bardeen-Zumino 
polynomial. 
We only need to compute the leading order in $\omega$.
For both cases $T=0$ and $T>0$ we can express the result as
\begin{align}
\sigma_{xy} &= 8\alpha A_z(r_0) \\
\sigma_{xx}=\sigma_{yy} &= \sqrt{h (r_0)}\,. 
\end{align}
For $T=0$ we have $r_0=0$ and $h(0)=0$ thus the diagonal conductivities 
vanish at zero temperature. 
The anomalous Hall effect (see Fig. \ref{fig:ahe}) is determined by the IR value 
of the axial gauge field. We can identify $b_\mathrm{eff} = A_z(r=0)$! 
We emphasize that this result is crucially dependent on the usage 
of the consistent (conserved) current. 
At zero temperature
it is non vanishing only in the second type of solutions described above. 
It therefore the topologically nontrivial solution with non-vanishing Hall 
conductivity. The third kind
of zero temperature solution is characterized by the restoration of 
time-reversal invariance at the end point of the holographic RG flow 
$A_z(0)=0$ and absence of Hall conductivity. 
Therefore we have a holographic model of a topological quantum phase transition 
between a topological and trivial semimetal. 
\vspace{.1cm}\\
\noindent{\bf Longitudinal conductivity:}
The longitudinal electric conductivity at both finite and zero temperature can 
be computed from the fluctuation $\delta V_z=v_z e^{-i\omega t}$ with equation 
of motion 
\begin{equation}
v_z''+\Big(\frac{f'}{f}-\frac{h'}{2h}+\frac{u'}{u}\Big)v_z'+\frac{\omega^2}{u^2}
v_z=0\,.
\end{equation}
At zero temperature we substitute $f=u$. 
We find
\begin{equation}\label{eq:sigmzz}
 \sigma_{zz} = \left.\frac{f}{\sqrt{h}}\right|_{r=r_0}\,.
\end{equation}

\begin{figure}
\begin{center}
\includegraphics[width=0.7\textwidth]{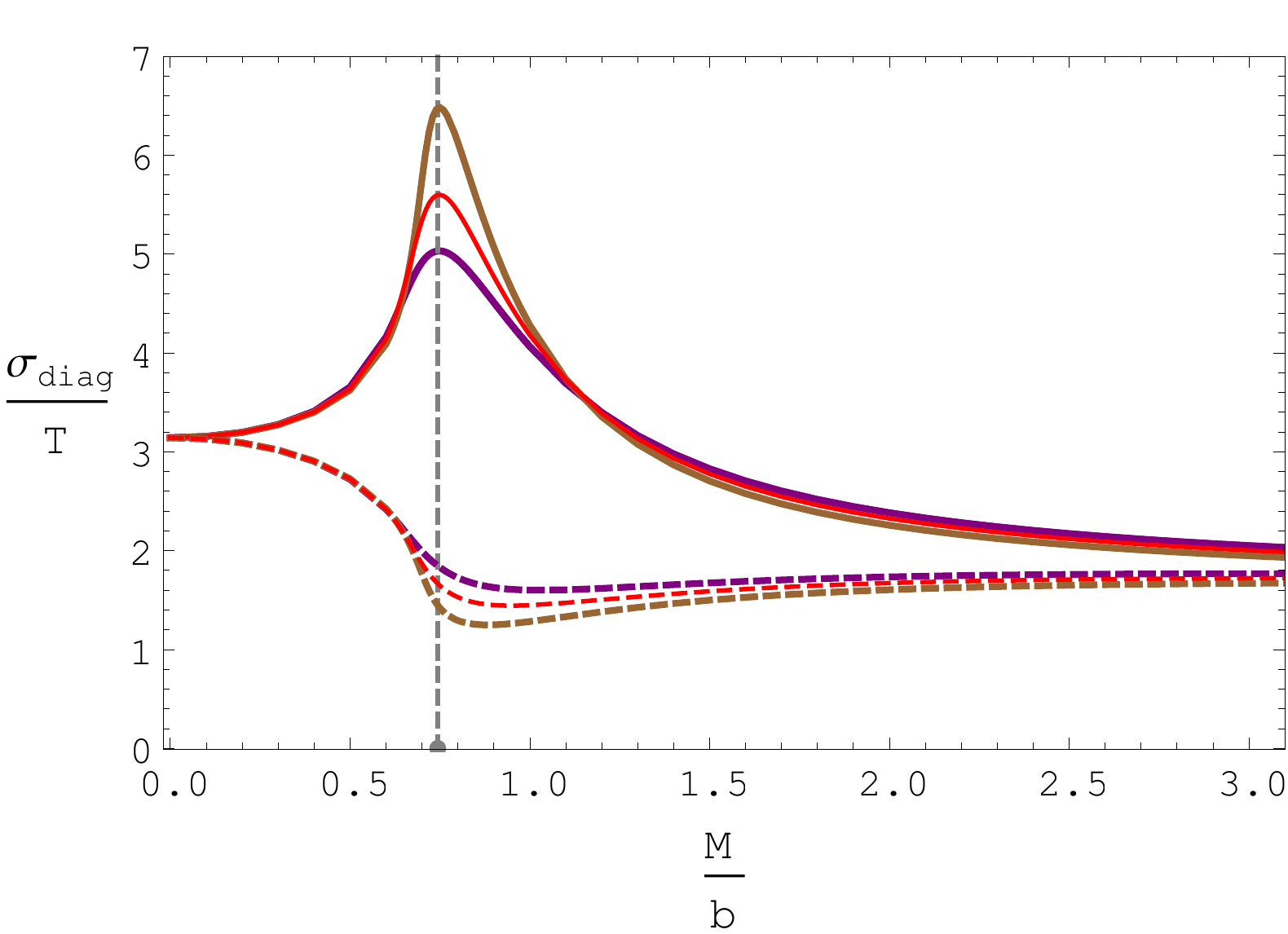}
\end{center}
\caption{\small The transverse and longitudinal electric conductivities for 
different temperatures. The solid lines are for $\sigma_{xx}=\sigma_{yy}$ and 
the dashed lines are for $\sigma_{zz}$ from our holographic model with $T/b= 
0.05$ (purple), $0.04$ (red), $0.03$ (brown). The dashed gray line is the 
critical value of $M/b$ at the topological phase transition.
}
\label{fig:con}
\end{figure}

The three types of background solutions can be classified according to the 
presence or absence of the anomalous Hall effect. 
There is a phase for $M/b$ smaller than a critical value 
in which the axial gauge field flows along the holographic direction towards a 
constant but nonzero value in the IR. The end point of this holographic flow of 
the axial gauge field determines the Hall conductivity $\sigma_{xy}$. 
At $M=0$ the flow is trivial and the Hall response is completely determined by 
the Chern-Simons current at
the boundary of AdS space (the Bardeen-Zumino polynomial). 
For $M\neq 0$ a nontrivial flow develops, the Hall 
conductivity has now two parts, a dynamical part, that can only be determined by 
solving
the equations (\ref{eq:flucsxy}) and the Chern-Simons part determined by the 
boundary values of the fields. At the critical value (for our choice of 
parameters this is $(M/b)_\mathrm{c} \simeq 0.744$) the Hall conductivity 
vanishes. At this value there is a critical
solution with a nontrivial scaling exponent in the $z$-direction. For even 
larger values of $M/b$ the solution shows no Hall effect. The axial gauge field 
flows to 
$A_z=0$ in the far IR. In contrast now the scalar field obtains a nontrivial IR 
value. This corresponds to the cosmological constant having a 
different value
in the far IR; i.e., the trivial solution is a domain wall in AdS similar to the 
zero temperature superconductor solutions described in Ref. 
\cite{Gubser:2009cg}.  Since in 
holography the cosmological constant is a measure of the effective number of 
degrees of freedom, the trivial solution can be interpreted as one in which some of 
the UV degrees of
freedom are gapped out along the RG flow. We have thus found a holographic zero 
temperature quantum phase transition between a topological phase characterized
by a non-vanishing Hall conductivity and a topological trivial phase with zero 
Hall conductivity. All diagonal conductivities vanish at zero temperature.

At $T\neq 0$ the quantum phase transition becomes a smooth crossover. 
The far IR physics is covered by a horizon at some finite value of  the
holographic coordinate. It is also interesting to observe the behavior of the 
diagonal conductivities at finite $T$ as a function of $M/b$ (see Fig. 
\ref{fig:con}). We see that the
transverse diagonal conductivities develop a peak roughly at the critical value 
whereas the longitudinal one develops a minimum. The height of the peak
and the depth of the minimum grow with temperature. At $M=0$ we simply have 
$\sigma_{xx,yy,zz}=\pi T$ and for large $M$ the conductivities
tend to a value $\sigma_{xx,yy,zz}= c \pi T$ with $c<1$ and independent of 
temperature. This is consistent with the interpretation that some but not all 
degrees of
freedom are gapped out in the trivial phase and that the phase transition is 
between a topological semimetal and a trivial semimetal. \\[1cm]

{\bf Hall viscosity}\\
So far we have shown that our holographic model of a Weyl semimetal has a 
quantum phase transition between a topological phase and a trivial phase 
distinguished by the presence 
of anomalous Hall conductivity. This is by itself interesting but so far the 
model does not do anything new compared to the free fermion model 
(\ref{eq:lorentzbreaking}).
So can we use the holographic model to compute something new? It turns out the 
answer is yes! In three dimensions anisotropic  time reversal breaking systems 
have 
a very complicated viscosity tensor. Viscosity can be defined as response to 
gradients in the fluid velocity
\begin{equation}
 T_{\mu  \nu} = -\eta_{\mu  \nu\rho\lambda} \partial_\rho u_\rho\,.
\end{equation}
Lorentz invariance restricts this to two independent components, the shear and 
the bulk viscosity (\ref{eq:hydroonederivativeT}). 
 More generally we can define the viscosities via the Kubo formula
\begin{equation}
\eta_{ij,kl}=\lim_{\omega\to 0}\frac{1}{\omega} 
\text{Im}\big[G^R_{ij,kl}(\omega,0)\big]\,,
\end{equation}
with the retarded Green's function of the energy momentum tensor
\begin{equation}
G^R_{ij,kl}(\omega, 0)=-\int dt d^3x e^{i\omega t} \theta(t) 
\langle[T_{ij}(t,{\vec x}), T_{kl}(0, 0)]\rangle\,.
\end{equation}
This has even and odd components under the exchange of the index pairs $ij$ and 
$kl$. We are interested in the
odd components. 
Choosing our coordinates such that ${\vec b} = b \hat e_z$, the two odd 
viscosities related to the anti-symmetric part of the 
retarded Green's function under the exchange of $(ij)\leftrightarrow(kl)$
are ( $T$ denotes here the index combination $xx-yy$)
\begin{equation}\label{eq:defoddvisoc}
\eta_{H_\parallel}=-\eta_{xz,yz}=\eta_{yz,xz}\,,~~~~\eta_{H_\perp}=\eta_{xy,T}
=-\eta_{T,xy}\,.
\end{equation}

These odd viscosity components can be calculated for the holographic
Weyl semi-metal. They are non-zero in the finite temperature 
backgrounds. The results can be expressed 
via the values of the bulk fields at the horizon \cite{Landsteiner:2016stv}
\begin{align}
\label{eq:oddvisperp}
\eta_{H_\parallel} &= 4 \zeta \frac{q^2 A_z \phi^2 f^2}{h} \bigg{|}_{r=r_0}\,,\\ 
\label{eq:oddvisparallel}
 \eta_{H_\perp} &=  8 \zeta q^2 \phi^2 f A_z \Big{|}_{r=r_0}\,.
\end{align}

\begin{figure}[h]s
\begin{center}
\includegraphics[width=0.49\textwidth]{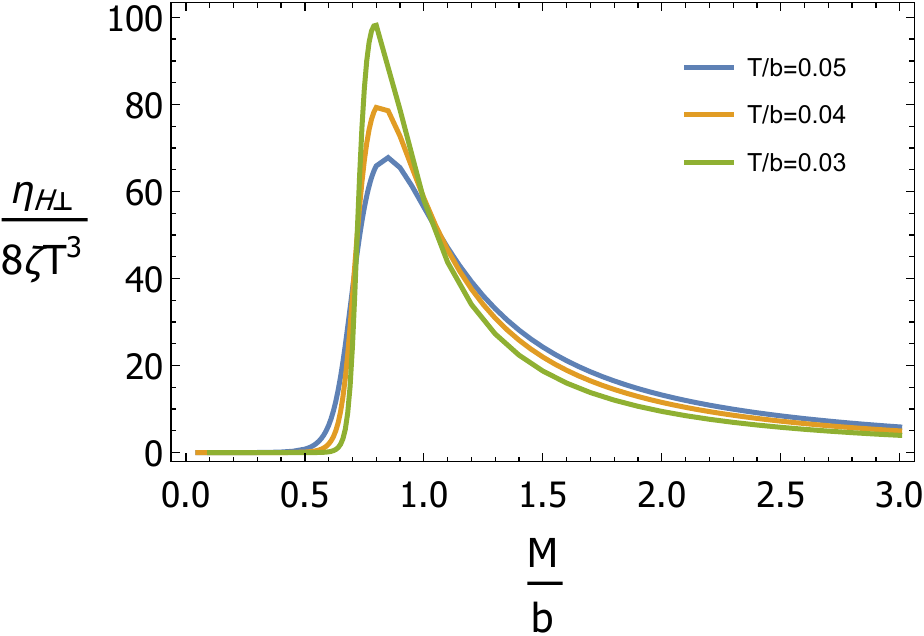}\hfill\includegraphics[
width=0.49\textwidth]{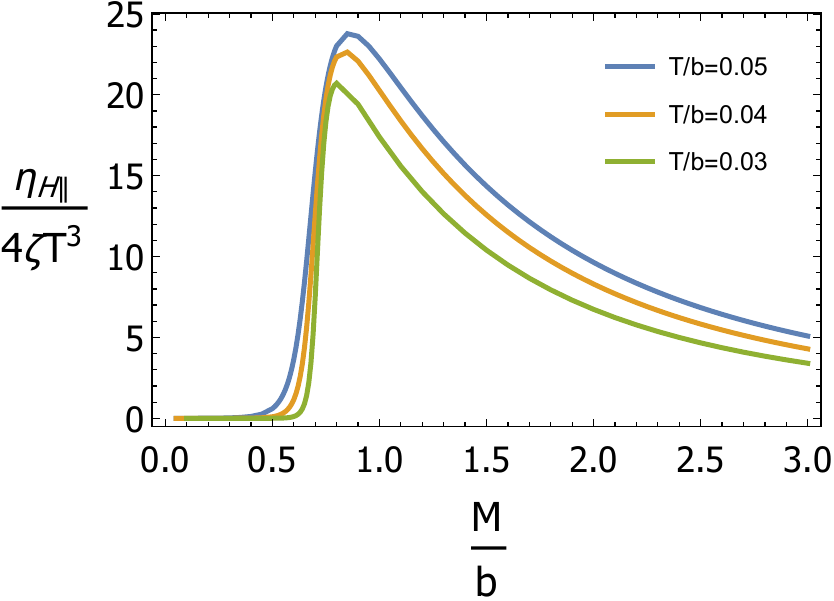}
\caption{\small Odd viscosity $\eta_{H_\perp}$(left panel) and $\eta_\parallel$  
as a function of $M/b$ at different low temperatures
normalized by $T^3$. }
\label{fig:oddvis_xyL}
\end{center}
\end{figure}

As can be seen from the plots the odd viscosities are very much suppressed in 
the topological phase. They rise steeply (in the chosen parametrization)
and peak near the critical value of $M/b$. Then they fall off again. This is an 
indication that the odd viscosity is a property related to to the underlying
quantum critical point that separates the topological from the trivial phase. In 
this region scaling laws for the odd viscosities (and other transport 
coefficients)
can be obtained from the expressions 
(\ref{eq:oddvisperp},\ref{eq:oddvisparallel}) \cite{Landsteiner:2016stv}. 
e.g. we find $\eta_{H\perp} \propto T^{2+\beta} $ and $\eta_{H,\parallel} \propto T^{4-\beta}$
and $\beta$ is the scaling exponent of $A_z$ of the critical solution.

\section{Outlook}
Anomalies are one of the cornerstones of quantum field theory. Almost 50 years 
after the realization that anomalies explain the decay of neutral pions they 
still are a major source of progress
theoretical and also experimental physics. In these notes I have summarized some 
of the story of anomalous transport phenomena emphasizing a few subtleties and 
hinting towards
some applications. While the basic phenomenology is now rather well understood, 
after a complicated history 
\cite{Vilenkin:1980fu, Vilenkin:1980zv, Volovik:2003fe, Alekseev:1998ds, Newman:2005as, 
Son:2004tq, Metlitski:2005pr} of discovery, neglect, re-discovery and final 
breakthrough there remain 
some pressing issues that need to be understood better.

First is the still somewhat mysterious way of how the mixed gravitational 
anomaly manages to influence transport at the one-derivative level. A hint is 
given by holography which allows
to swallow surplus derivatives up in the extra dimension. Steps towards a 
holography independent understanding have been made, e.g. combining 
hydrodynamics with geometric arguments \cite{Jensen:2012kj},
non-renormalization theorems \cite{Golkar:2012kb,Hou:2012xg}, 
considering Berry flux through Fermi surfaces \cite{Basar:2013qia}  
and links to global gravitational anomalies \cite{Golkar:2015oxw}.  
Beyond that there is a pressing need of addressing the experimental side of the 
gravitational anomaly. In high energy physics direct measurement of pion decay 
into gravitons seems
hopeless but in condensed matter be it the quark gluon plasma or the electron 
fluid of Weyl semi-metals the collective transport phenomena induced by the 
gravitational anomaly 
are in principle accessible. Hopefully ingenious experimental physicists will 
get excited about this possibility in the near future. 

While holography can probably not claim to have discovered  anomalous transport 
it has certainly played a major role in gaining a better understanding. But the 
holographic story has not yet
ended: as we have reviewed a holographic model of a Weyl semimetal state shows 
very unusual viscosity properties in the quantum critical region that lies 
between the topological
and trivial phase. Viscous flow of the electron fluid in Graphene has recently 
been measured \cite{Electronvis1, Electronvis2, Electronvis3}. 
So one naturally hopes that this (string theory based prediction) 
of odd viscosity in 
the quantum critical region of Weyl semi-metals can be measured one day as well. 

There 
are man aspects that are missing from this review. Especially the application of 
anomalous transport theory
to the physics of the quark gluon plasma. Suffice it to point to the recent 
reviews \cite{Kharzeev:2015znc, Skokov:2016yrj, Huang:2015oca}. 
Another important subject totally missing from these lectures is chiral kinetic theory
\cite{Son:2012wh, Stephanov:2012ki, Manuel:2014dza, Dwivedi:2013dea}. 
Kubo formulas for anomalies transport have been introduced and studied in \cite{Kharzeev:2009pj,
Amado:2011zx,
Landsteiner:2011cp, Landsteiner:2013aba, Chowdhury:2015pba}. This approach is especially well suited
to study the frequency dependence of anomalous transport coefficients. Another very 
systematic approach to anomalous transport has been developed 
in \cite{Loganayagam:2012zg}.
Anomalies in $d$ dimensions are governed via the so-called descent equations
by invariant polynomials in the field strengths in $d+2$ dimensions. 
The anomalous
currents can be obtained from the invariant polynomials substitution rule $F\rightarrow \mu$,
$(p_2(R)\rightarrow -T^2, p_{k>1}\rightarrow 0)$ where $p_k(R)$ is the Pontryagin
classes, i.e. invariant polynomials in the Riemmann tensor, $p_2(R)$ is the
gravitational contribution to the chiral anomaly. 

We have always assumed that the electric
and magnetic fields are external and non-dynamical. Including dynamics of the 
gauge fields is however an important issue and leads to several new aspects. 
First of all the
the actual axial QCD axial anomaly has a contribution form the gluon fields 
which have strong quantum dynamics in physically interesting situations such as 
heavy ion collisions. This allows
processes that actually create net chiral charge and was the origin of the idea 
of the presence of the chiral magnetic effect in heavy ion collisions 
\cite{Kharzeev:2007jp}. Also the values for the anomalous transport coefficients
are affected \cite{Hou:2012xg, Jensen:2013vta}
In holography anomalies with dynamical gluons can be modelled by using the 
St{\"u}ckelberg mechanism in the bulk of AdS \cite{Gursoy:2014boa,Jimenez-Alba:2014iia}. 
Coupling the chiral magnetic current to Maxwells
equations leads to the so-called chiral magnetic instability 
\cite{Akamatsu:2013pjd,Manuel:2015zpa,Hirono:2015rla,Gorbar:2016klv} converting axial 
chemical potentials into helical magnetic fields.

\section{Acknowledgement}
These are the notes of lectures held at the 56th Cracow School on Theoretical 
Physics  in Zakopane, Poland and the Focus workshop on Holography and Topology of Quantum Matter 
in Pohang, Korea.
I would like to thank the organizers of both.
I also thank all my collaborators for helping me to unravel and understand some 
of the issues involved.
My research has been supported by  FPA2015-65480-P and by the Centro de 
Excelencia Severo Ochoa Programme under grant SEV-2012-0249.

\bibliography{AnomTrans}{}
\bibliographystyle{hunsrt}
\end{document}